\documentclass[aps,preprint,floats,epsf,epsfig,nofootinbib,letter]{revtex4}

\usepackage{graphicx}
\usepackage{dcolumn}
\usepackage{bm}

\def\ol{\overline}
\def\ov{\overline}
\def\be{\begin{eqnarray}}
\def\en{\end{eqnarray}}
\def\non{\nonumber}
\def\la{\langle}
\def\ra{\rangle}
\def\A{{\cal A}}
\def\B{{\cal B}}
\begin{document}

\renewcommand{\baselinestretch}{1.20}

\centerline{\Large\bf Two-body hadronic charmed meson decays}

\bigskip
\centerline{\bf Hai-Yang Cheng,$^{1,2}$ Cheng-Wei Chiang$^{3,1}$ }
\medskip
\centerline{$^1$ Institute of Physics, Academia Sinica}
\centerline{Taipei, Taiwan 115, Republic of China}
\medskip
\centerline{$^2$ Physics Department, Brookhaven National
Laboratory} \centerline{Upton, New York 11973}
\medskip
\centerline{$^3$ Department of Physics and Center for Mathematics } \centerline{and Theoretical Physics, National Central University}
\centerline{Chung-Li, Taiwan 320, Republic of China}
\medskip

\centerline{\bf Abstract}
\bigskip
\small

We study in this work the two-body hadronic charmed meson decays, including both the $PP$ and $VP$ modes.  The latest experimental data are first analyzed in the diagrammatic approach.  The magnitudes and strong phases of the flavor amplitudes are extracted from the Cabibbo-favored (CF) decay modes using $\chi^2$ minimization.  The best-fitted values are then used to predict the branching fractions of the singly-Cabibbo-suppressed (SCS) and doubly-Cabibbo-suppressed decay modes in the flavor SU(3) symmetry limit.  We observe significant SU(3) breaking effects in some of SCS channels.  In the case of $VP$ modes, we point out that the $A_P$ and $A_V$ amplitudes cannot be completely determined based on currently available data.  We conjecture that the quoted experimental results for both $D_s^+\to\bar K^0K^{*+}$ and $D_s^+\to \rho^+\eta'$ are overestimated. We compare the sizes of color-allowed and color-suppressed tree amplitudes extracted from the diagrammatical approach with the effective parameters $a_1$ and $a_2$ defined in the factorization approach.  The ratio $|a_2/a_1|$ is more or less universal among the $D \to {\bar K} \pi$, ${\bar K}^* \pi$ and ${\bar K} \rho$ modes.  This feature allows us to discriminate between different solutions of topological amplitudes.  For the long-standing puzzle about the ratio $\Gamma(D^0\to K^+K^-)/\Gamma(D^0\to\pi^+\pi^-)$, we argue that, in addition to the SU(3) breaking effect in the spectator amplitudes, the long-distance resonant contribution through the nearby resonance $f_0(1710)$ can naturally explain why $D^0$ decays more copiously to $K^+ K^-$ than $\pi^+ \pi^-$ through the $W$-exchange topology. This has to do with the dominance of the scalar glueball content of $f_0(1710)$ and the chiral-suppression effect in the decay of a scalar glueball into two pseudoscalar mesons. The same FSI also explains the occurrence of $D^0\to K^0\bar K^0$ and its vanishing amplitude when SU(3) flavor symmetry is exact.  Owing to the $G$-parity selection rule, $D_s^+\to\pi^+\omega$ does not receive contributions from the short-distance $W$-annihilation and resonant FSIs, but it can proceed through the weak decays $D_s^+\to\rho^+\eta^{(')}$ followed by the final-state rescattering of $\rho^+\eta^{(')}$ into $\pi^+\omega$ through quark exchange.

\pacs{Valid PACS appear here}

\maketitle

%
%
\section{Introduction \label{sec:intro}}

The hadronic decays of charmed mesons and related physics have been studied extensively in the past 35 years since the first discovery of charmed meson states in 1974. The experimental progress is still impressive. For example, many hadronic multibody decays of $D^0,D^+$ and $D_s^+$ mesons have been studied using the technique of Dalitz plot analysis in which the resonant structure is probed. The Dalitz analysis offers the opportunity to understand the light scalar meson spectroscopy and the underlying structure of light scalar mesons $\sigma, \kappa, a_0(980)$ and $f_0(980)$.  The study of charm physics also opens a new avenue to the search of new physics.  For example, the recent observation of $D^0-\bar D^0$ mixing will enable us to explore new physics via flavor-changing neutral currents in the {\bf up}-type quark sector.  The standard model predicts almost null {\it CP} asymmetries in the charm system.  Hence, any observation of {\it CP} violation in hadronic charm decays will signal physics beyond the standard model.

With the advent of heavy quark effective theory, it is known that physics of nonleptonic $B$ decays can be formulated in a QCD-inspired approach such as QCD factorization, pQCD and soft-collinear effective theory.  However, until today a theoretical description of the underlying mechanism for exclusive hadronic $D$ decays based on QCD is still not yet available.  This has to do with the mass of the charm quark, of order 1.5 GeV.  It is not heavy enough to allow for a sensible heavy quark expansion and it is not light enough for the application of chiral perturbation theory.  It is known that the na{\"i}ve factorization approach fails to describe color-suppressed (class-II) decay modes. Empirically, it was learned in the 1980s that if the Fierz-transformed terms characterized by $1/N_c$ are dropped, the discrepancy between theory and experiment will be greatly improved \cite{Fuk}.  This leads to the so-called large-$N_c$ approach for hadronic $D$ decays \cite{Buras}.  Theoretically, explicit calculations based on the QCD sum-rule analysis \cite{BS} indicate that the Fierz terms are indeed largely compensated by the nonfactorizable corrections. However, the $1/N_c$ ansatz is not applicable to $B$ decays.

In spite of the difficulties in formulating a theoretical framework for describing nonleptonic decays of charmed mesons, a model-independent analysis of charm decays based on the diagrammatic approach is possible. In this approach, the flavor-flow diagrams are classified according to the topologies of weak interactions with all strong interaction effects included.  Based on flavor SU(3) symmetry, this model-independent analysis enables us to extract the topological amplitudes and sense the relative importance of different underlying decay mechanisms.  When enough measurements are made with sufficient accuracy, we can extract the diagrammatic amplitudes from experiment and compare to theoretical estimates, especially checking whether there are any significant final-state interactions or whether the weak annihilation diagrams can be ignored as often asserted in the literature.

A salient feature of the diagrammatic approach analysis of charm decays is that, unlike the $B$ decays, the magnitude and relative strong phase of each individual topological diagram can be extracted from the data without making further assumptions apart from flavor SU(3) symmetry.  Explicitly, the color-allowed amplitude $T$, the color-suppressed amplitude $C$, $W$-exchange amplitude $E$ and $W$-annihilation amplitude $A$ in Cabibbo-allowed $D\to PP$ decays can be cleanly extracted from the data.  Recall that the ratio $|C/T|$ in charmless $B$ decays cannot be unambiguously determined due to the contamination of  electroweak penguin or weak annihilation diagrams and that weak annihilation topologies in the $B$ sector are usually neglected in the diagrammatic approach analysis.

The topological diagram analysis of $D$ decays evidently shows the importance of weak annihilation ($W$-exchange or $W$-annihilation) contributions.  Since the short-distance weak annihilation amplitude is subject to helicity suppression, this implies that long-distance weak annihilation induced by final-state interactions should play an essential role.

In this work we shall study the two-body hadronic charmed meson decays, including both the $PP$ and $VP$ modes.  We first analyze the latest experimental data in the diagrammatic approach.  For $D\to PP$ decays, there are new CLEO measurements with better precision \cite{CLEOPP09}.  There are two different ways of extracting the topological amplitudes for $D\to VP$ decays.  We shall present both results and make a comparison between them.  At present, there exist many possible solutions for the flavor amplitudes.  We will study the effective parameters $a_1$ and $a_2$ defined in the factorization approach and use them as a means to discriminate among different solutions.  Another goal of this work is to see if we can understand the SU(3) breaking effects occurring in the SCS modes, in the hope of shedding more light on or even resolving the long-standing puzzle with the ratio of $D^0\to K^+K^-$ to $D^0\to \pi^+\pi^-$ branching fractions.

The layout of the present paper is as follows.  After a brief review of the diagrammatic approach, we extract magnitudes and strong phases of the flavor amplitudes from CF $D\to PP$ and $D\to VP$ decays in Sec.~II.  From the extracted invariant amplitudes, we determine the effective parameters $a_1$ and $a_2$ defined in the factorization approach in Sec.~III and proceed to discuss SU(3) breaking effects in singly-Cabibbo-suppressed modes.  Using the branching fraction ratio of $D^0\to K^+K^-$ to $D^0\to \pi^+\pi^-$ as an example, we illustrate how to explain the large SU(3) violation in the ratio by considering the $W$-exchange topology induced by the nearby resonance. Sec.~IV comes to our conclusions.

For nice and comprehensive reviews of exclusive hadronic charmed meson decays, the reader is referred to Refs.~\cite{Ryd,Wu}.

\section{Diagrammatic approach \label{sec:topology}}

\subsection{Preamble}

It has been established sometime ago that a least model-dependent analysis of heavy meson decays can be carried out in the so-called quark-diagram approach \footnote{It is also referred to as flavor-diagram or topological-diagram approach in the literature.}. In this diagrammatic scenario, all two-body nonleptonic weak decays of heavy mesons can be expressed in terms of six distinct quark diagrams \cite{Chau,CC86,CC87}:\footnote{Historically, the quark-graph amplitudes $T,\,C,\,E,\,A,\,P$ named in \cite{Gronau94} were originally denoted by $A,\,B,\,C,\,D,\,E$, respectively, in \cite{CC86,CC87}.  For the analysis of charmless $B$ decays, one adds the variants of the penguin diagram such as the electroweak penguin and the penguin annihilation.} $T$, the color-allowed external $W$-emission tree diagram; $C$, the color-suppressed internal $W$-emission diagram; $E$, the $W$-exchange diagram; $A$, the $W$-annihilation diagram; $P$, the horizontal $W$-loop diagram; and $V$, the vertical $W$-loop diagram.  (The one-gluon exchange approximation of the $P$ graph is the so-called ``penguin diagram''.)  It should be stressed that these diagrams are classified according to the topologies of weak interactions with all strong interaction effects encoded, and hence they are {\it not} Feynman graphs.  All quark graphs used in this approach are topological and meant to have all the strong interactions included, {\it i.e.}, gluon lines are included implicitly in all possible ways.  Therefore, analyses of topological graphs can provide information on final-state interactions (FSIs).  Various topological amplitudes in two-body hadronic $D$ decays have been extracted from the data in \cite{Rosner99,Chiang03,Wu04,Wu05,RosnerPP08,RosnerVP,CLEOPP08,RosnerPP09} after making some reasonable approximations, {\it e.g.}, flavor SU(3) symmetry.

For charm decays involving an SU(3) singlet in the final product, {\it e.g.}, $D^0\to \ov K^0\phi,~\ov K^0\omega,~\ov K^0\eta_0$, there exist additional hairpin diagrams in which a quark-antiquark pair is created from vacuum to form a color- and flavor-singlet final-state meson \cite{CC89,Li}.  There are four different types of disconnected hairpin diagrams: $E_h,\,A_h,\,P_h,\,V_h$ corresponding to the topological graphs $E,\,A,\,P,\,V$ (for details, see \cite{CC89}).  The amplitudes $E_h$ and $A_h$ are denoted by $SE$ and $SA$, respectively, in Ref.~\cite{Chiang03}.  They appear in final states with $\eta^{(\prime)}$ for the $PP$ modes and with $\phi$, $\omega$, and/or $\eta^{(\prime)}$ for the $PV$ modes.  A test of their importance has been proposed in \cite{Chiang03}.  As they are still not strongly called for phenomenologically (see discussions in Section~\ref{subsec:DPP}), we will ignore these diagrams in the current analysis.  For hadronic charm decays, contributions from the diagrams $P$ and $V$ are negligible due to the cancellation of the CKM matrix elements $V_{cd}^*V_{ud}$ and $V_{cs}^*V_{us}$.

For final states involving an $\eta$ or $\eta'$, it is more convenient to consider the flavor mixing of $\eta_q$ and $\eta_s$ defined by
 \be
 \eta_q={1\over\sqrt{2}}(u\bar u+d\bar d),\qquad\quad
 \eta_s=s\bar s,
 \en
in analogy with the wave functions of $\omega$ and $\phi$ in ideal mixing.  The wave functions of the $\eta$ and $\eta'$ are given by
 \be
 \left(\matrix{ \eta \cr \eta'\cr}\right)=\left(\matrix{ \cos\phi & -\sin\phi \cr
 \sin\phi & \cos\phi\cr}\right)\left(\matrix{\eta_q \cr \eta_s
 \cr}\right),
 \en
where $\phi=\theta+{\rm arctan}\sqrt{2}$, and $\theta$ is the $\eta\!-\!\eta'$ mixing angle in the octet-singlet basis
 \be
 \left(\matrix{ \eta \cr \eta'\cr}\right)=\left(\matrix{ \cos\theta & -\sin\theta \cr
 \sin\theta & \cos\theta\cr}\right)\left(\matrix{\eta_8 \cr \eta_0
 \cr}\right).
 \en
The most recent experimental determination of the $\eta\!-\!\eta'$ mixing angle is $\phi=(40.4\pm 0.6)^\circ$ from KLOE \cite{KLOE}, \footnote{For the mixing angle $\phi={\rm arc\,sin}(1/\sqrt{3})=35.26^\circ$, the $\eta$ and $\eta'$ wave functions have simple expressions \cite{Chau91}: $ \eta = {1\over \sqrt{3}}(\sqrt{2}\eta_q-\eta_s)={1\over
 \sqrt{3}}(u\bar u+d\bar d-s\bar s)$ and
$ \eta'= {1\over\sqrt{3}}(\eta_q+\sqrt{2}\eta_s)={1\over
 \sqrt{6}}(u\bar u+d\bar d+2s\bar s)$.}
which is indeed close to the original theoretical and phenomenological estimates of $42.2^\circ$ and $(39.3\pm1.0)^\circ$, respectively, made by Feldmann, Kroll and Stech \cite{FKS}.  Notice that charm decays involving a final-state $\eta_s$ are governed by only one type of topological amplitudes, in addition to the associated hairpin diagram.  For example,
\be
\A(D^0\to \bar K^0\eta_s)=E, \qquad && \A(D^0\to \bar K^{*0}\eta_s)=E_V,  \non \\
\A(D_s^+\to \pi^+\eta_s)=T, \qquad && \A(D_s^+\to \rho^+\eta_s)=T_P.
\en
The $D\to M\eta$ and $M\eta'$ amplitudes then have the expressions
 \be \label{etaetap}
 \A(D\to M\eta) &=& \A(D\to M\eta_q)\cos\phi-\A(D\to
 M\eta_s)\sin\phi, \non \\
 \A(D\to M\eta') &=& \A(D\to M\eta_q)\sin\phi+\A(D\to
 M\eta_s)\cos\phi.
 \en

\subsection{$D\to PP$ \label{subsec:DPP}}

Based on flavor SU(3) symmetry, the quark-graph amplitudes for the Cabibbo-favored (CF), singly-Cabibbo-suppressed (SCS) and doubly-Cabibbo-suppressed (DCS) decays of charmed mesons into two pseudoscalar mesons are listed in Tables \ref{tab:CFPP} to \ref{tab:DCSPP}, respectively.  Experimental data on the branching fractions are taken from the most recent CLEO measurements \cite{CLEOPP09}.  We shall follow the conventional practice to denote the primed amplitudes for SCS modes and double-primed amplitudes for DCS decays. In the SU(3) limit, primed and unprimed amplitudes should be the same.  Note that the selection rule for a vanishing $D_s^+\to\pi^+\pi^0$ follows from the isospin transformation properties of the weak Hamiltonian and isospin invariance of strong interactions and hence it is unaffected by SU(3) breaking or final-state interactions \cite{Lipkin80}.

For $D\to PP$ decays, the amplitudes $T,C,E,A$ have dimensions of energy as they are related to the partial decay rate via
 \be
 \Gamma(D\to PP)={p_c\over 8\pi m_D^2}|\A|^2,
 \en
with $p_c$ being the center-of-mass momentum of either meson in the final state.  The reduced quark-graph amplitudes $T,C,E,A$ are extracted from the Cabibbo-allowed $D\to PP$ decays to be (in units of $10^{-6}$ GeV)
\be \label{eq:PP1}
&& T=3.14\pm0.06, \qquad\qquad\qquad\quad
C=(2.61\pm0.08)\,e^{-i(152\pm1)^\circ}, \non \\
&&  E=(1.53^{+0.07}_{-0.08})\,e^{i(122\pm2)^\circ},
\qquad\quad  A=(0.39^{+0.13}_{-0.09})\,e^{i(31^{+20}_{-33})^\circ}
\en
for $\phi=40.4^\circ$, and
\be \label{eq:PP2}
&& T=3.08\pm0.06\,, \qquad\qquad\qquad\quad
C=(2.46^{+0.06}_{-0.07})\,e^{-i(152\pm1)^\circ}, \non \\
&&  E=(1.66\pm0.06)\,e^{i(120\pm2)^\circ},
\qquad A=(0.34^{+0.17}_{-0.18})\,e^{i(70^{+10}_{-27})^\circ}
\en
for $\phi=35.26^\circ$.  The fitted $\chi^2$ value is 0.29 per degree of freedom with quality $59.2\%$ and 1.73 per degree of freedom with quality $18.8\%$ for the fits (\ref{eq:PP1}) and (\ref{eq:PP2}), respectively.  Throughout this paper, we take $\lambda = 0.2253 \pm 0.0007$, the weighted average of CKMfitter \cite{CKMfitter} and UTfit \cite{UTfit} groups, and neglect its small error.  The relevant CKM matrix elements are $V_{ud} = V_{cs} = 1 - \lambda^2/2$ and $V_{us} = - V_{cd} = \lambda$.  With the inclusion of CKM factor $V_{cs}^*V_{ud}$ in the amplitudes, our results in (\ref{eq:PP2}) agree with those in \cite{RosnerPP09}.  Note that the previous fit in 2008 had \cite{RosnerPP08,CLEOPP08}
\be \label{eq:PPold2}
&& T=2.78\pm0.13, \qquad\qquad\qquad\quad
C=(2.04\pm0.17)\,e^{i(-151\pm2)^\circ}, \non \\
&&  E=(1.68\pm0.12)\,e^{i(117\pm4)^\circ},
\qquad A=(0.54\pm0.37)\,e^{i(-64^{+32}_{-~8})^\circ}
\en
with $\chi^2=0.65$ per degree of freedom.

A few words about error extraction in our analysis are in order here.  To obtain the 1-$\sigma$ range of some parameter around its best-fitted value, we scan the parameter while allowing all the other parameters to vary at the same time to minimize the $\chi^2$ function.  This renders a more conservative range than the method that fixes all the other parameters to their best-fitted values when scanning the parameter of interest, as used, for example, in Ref.~\cite{RosnerVP}.

The magnitudes of the updated $T$ and $C$ are larger than the 2008 ones due to the newly measured branching fractions of $\bar K^0\eta$, $\pi^+\eta$ and $\pi^+\eta'$ modes being larger than the old ones.  Notice that the $W$-annihilation amplitude $A$ is substantially smaller than the $W$-exchange one
$E$, and it is no longer antiparallel to $E$, contrary to the previous claim \cite{Rosner99,RosnerPP08}. Indeed, $A$ is almost perpendicular to $E$ in fit (\ref{eq:PP1}) and its angle relative to $E$ is close to $50^\circ$ in fit (\ref{eq:PP2}).

\begin{table}[t]
\caption{Branching fractions and invariant amplitudes for Cabibbo-favored decays of charmed mesons to two pseudoscalar mesons.  Data are taken from \cite{CLEOPP09}.  Predictions based on our best-fitted results in (\ref{eq:PP1}) are given in the last column.
  \label{tab:CFPP}}
\vspace{6pt}
\begin{ruledtabular}
\begin{tabular}{l l l c c c c}
Meson & Mode & Representation
     & ${\cal B}_{\rm exp}$ &  ${\cal B}_{\rm fit}$ \\
 & & & ($\%$)  & ($\%$) \\
\hline
$D^0$ & $K^{-} \pi^+$ & $V^*_{cs}V_{ud}(T + E)$
     & $3.91 \pm 0.08$   & $3.91 \pm 0.17$ \\
 & $\ol{K}^{0} \pi^0$ & $\frac{1}{\sqrt{2}}V^*_{cs}V_{ud}(C - E)$
     & $2.38 \pm 0.09$   & $2.36 \pm 0.08$ \\
 & $\ol{K}^{0} \eta$ & $V^*_{cs}V_{ud}[{1\over \sqrt{2}}(C + E)\cos\phi - E\sin\phi\,]$
     & $0.96 \pm 0.06$ & $0.98 \pm 0.05$ \\
 & $\ol{K}^{0} \eta\,'$ & $V^*_{cs}V_{ud}[{1\over \sqrt{2}}(C + E)\sin\phi + E\cos\phi\,]$
     & $1.90\pm0.11$ & $1.91 \pm 0.09$ \\
\hline
$D^+$ & $\ol{K}^{0} \pi^+$ & $V^*_{cs}V_{ud}(T + C)$
     & $3.07 \pm 0.10$  & $3.08 \pm 0.36$ \\
\hline
$D_s^+$ & $\ol{K}^{0} K^+$ & $V^*_{cs}V_{ud}(C + A)$
     & $2.98 \pm 0.17$  & $2.97 \pm 0.32$ \\
 & $\pi^+ \pi^0$ & $0$
     & $<0.037$  & $0$ \\
 & $\pi^+ \eta$ & $V^*_{cs}V_{ud}(\sqrt{2}A\cos\phi - T\sin\phi)$
     & $1.84 \pm 0.15$  & $1.82 \pm 0.32$ \\
 & $\pi^+ \eta\,'$ & $V^*_{cs}V_{ud}(\sqrt{2}A\sin\phi + T\cos\phi)$
     & $3.95 \pm 0.34$  & $3.82 \pm 0.36$ \\
\end{tabular}
\end{ruledtabular}
\end{table}

\begin{table}
\caption{Branching fractions and invariant amplitudes for singly-Cabibbo-suppressed decays of charmed mesons to two pseudoscalar mesons. Data are taken from \cite{CLEOPP09}.  Predictions based on our best-fitted results in (\ref{eq:PP1}) with exact flavor SU(3) symmetry are given in the last column.
  \label{tab:CSPP}}
\begin{ruledtabular}
\begin{tabular}{l l l c c c c}
Meson & Mode & Representation
     & ${\cal B}_{\rm exp}$ & ${\cal B}_{\rm theory}$ \\
 & & & ($\times 10^{-3}$)  & ($\times 10^{-3}$) \\
 \hline
$D^0$
  & $\pi^+ \pi^-$ & $V^*_{cd}V_{ud}(T'+E')$
     & $1.45 \pm 0.05$ & $2.24 \pm 0.10$ \\
  & $\pi^0 \pi^0$ & ${1\over\sqrt{2}}V^*_{cd}V_{ud}(C'-E')$
     & $0.81 \pm 0.05$  & $1.35 \pm 0.05$ \\
  & $\pi^0 \eta $ & $-V^*_{cd}V_{ud}E'\cos\phi-{1\over\sqrt{2}}V^*_{cs}V_{us}C'\sin\phi$
     & $0.68 \pm 0.07$ & $0.75 \pm 0.02$ \\
  & $\pi^0 \eta' $ & $-V^*_{cd}V_{ud}E'\sin\phi+{1\over\sqrt{2}}V^*_{cs}V_{us}C'\cos\phi$
     & $0.91 \pm 0.13$  & $0.74 \pm 0.02$ \\
  & $\eta\eta $ & $-{1\over\sqrt{2}}V^*_{cd}V_{ud}(C'+E')\cos^2\phi+V^*_{cs}V_{us}
  (2E'\sin^2\phi-{1\over\sqrt{2}}C'\sin 2\phi)$
     & $1.67 \pm 0.18$ & $1.44 \pm 0.08$ \\
  & $\eta\eta' $ & $-{1\over 2}V^*_{cd}V_{ud}(C'+E')\sin 2\phi+V^*_{cs}V_{us}(E'\sin 2\phi-{1\over\sqrt{2}}C'\cos 2\phi)$
     & $1.05 \pm 0.26$  & $1.19 \pm 0.07$ \\
  & $K^+ K^{-}$ & $V^*_{cs}V_{us}(T'+E')$
     & $4.07 \pm 0.10$  & $1.92 \pm 0.08$ \\
  & $K^0 \ol{K}^{0}$ & $V^*_{cd}V_{ud}E'_s+V^*_{cs}V_{us}E'_d$ \footnotemark[1]
     & $0.64\pm0.08$  & $0$ \\
\hline
$D^+$
  & $\pi^+ \pi^0$ & $\frac{1}{\sqrt{2}}V^*_{cd}V_{ud}(T'+C')$
     & $1.18 \pm 0.07$  & $0.88 \pm 0.10$ \\
  & $\pi^+ \eta $ & $\frac{1}{\sqrt{2}}V^*_{cd}V_{ud}(T'+C'+2A')\cos\phi-V^*_{cs}V_{us}C'\sin\phi$
     & $3.54 \pm 0.21$ & $1.48 \pm 0.26$ \\
  & $\pi^+ \eta' $ & $\frac{1}{\sqrt{2}}V^*_{cd}V_{ud}(T'+C'+2A')\sin\phi+V^*_{cs}V_{us}C'\cos\phi$
     & $4.68 \pm 0.30$ & $3.70 \pm 0.37$ \\
  & $K^+ \ol{K}^{0}$ & $V^*_{cd}V_{ud}A'+V^*_{cs}V_{us}T'$
     & $6.12 \pm 0.22$ \footnotemark[2]  & $5.46 \pm 0.53$ \\
\hline
$D_s^+$
  & $\pi^+ K^{0}$ & $V^*_{cd}V_{ud}T'+V^*_{cs}V_{us}A'$
     & $2.52 \pm 0.27$ \footnotemark[3]  & $2.73 \pm 0.26$ \\
  & $\pi^0 K^{+}$ & $\frac{1}{\sqrt{2}}(V^*_{cd}V_{ud}C'-V^*_{cs}V_{us}A')$
     & $0.62\pm 0.23$  & $0.86 \pm 0.09$  \\
  & $K^{+}\eta$ & $\frac{1}{\sqrt{2}}(V^*_{cd}V_{ud} C'+V^*_{cs}V_{us} A')\cos\phi-V^*_{cs}V_{us}(T'+C'+A')\sin\phi$
  & $1.76\pm0.36$  & $0.78 \pm 0.09$ \\
  & $K^{+}\eta' $ & $\frac{1}{\sqrt{2}}(V^*_{cd}V_{ud} C'+V^*_{cs}V_{us} A')\sin\phi+V^*_{cs}V_{us}(T'+C'+A')\cos\phi$
  & $1.80\pm 0.52$  & $1.07 \pm 0.17$ \\
\end{tabular}
\footnotetext[1]{The subscript $q$ in $E'_q$ refers to the quark-antiquark pair popping out of the vacuum in the final state.}
\footnotetext[2]{A new Belle measurement yields $\B(D^+\to K^+\bar K^0)=(5.50\pm0.16)\times 10^{-3}$ \cite{BelleDs}. }
\footnotetext[3]{A new Belle measurement yields $\B(D_s^+\to \pi^+K^0)=(2.40\pm0.18)\times 10^{-3}$ \cite{BelleDs}. }
\end{ruledtabular}
\end{table}
\begin{table}[t]
\caption{Branching fractions and invariant amplitudes for doubly-Cabibbo-suppressed decays of charmed mesons to two pseudoscalar mesons. Data are taken from \cite{CLEOPP09}.  Predictions based on our best-fitted results in (\ref{eq:PP1}) with exact flavor SU(3) symmetry are given in the last column.
  \label{tab:DCSPP}}
\vspace{6pt}
\begin{ruledtabular}
\begin{tabular}{l l l c c c c}
Meson & Mode & Representation
     & ${\cal B}_{\rm exp}$  & ${\cal B}_{\rm theory}$ \\
 & & & ($\times 10^{-4}$)  & ($\times 10^{-4}$) \\
\hline
$D^0$ & $K^{+} \pi^-$ & $V^*_{cd}V_{us}(T'' + E'')$
     & $1.48 \pm 0.07$  & $1.12 \pm 0.05$ \\
 & ${K}^{0} \pi^0$ & $\frac{1}{\sqrt{2}}V^*_{cd}V_{us}(C'' - E'')$
     &   & $0.67 \pm 0.02$ \\
 & ${K}^{0} \eta$ & $V^*_{cd}V_{us}[{1\over \sqrt{2}}(C'' + E'')\cos\phi - E''\sin\phi\,]$
     &  & $0.28 \pm 0.02$ \\
 & ${K}^{0} \eta\,'$ & $V^*_{cd}V_{us}[{1\over \sqrt{2}}(C'' + E'')\sin\phi + E''\cos\phi\,]$
     &  & $0.55 \pm 0.03$ \\
\hline
$D^+$ & ${K}^{0} \pi^+$ & $V^*_{cd}V_{us}(C''+A'')$
     &  & $1.98 \pm 0.22$ \\
     & ${K}^{+} \pi^0$ & ${1\over\sqrt{2}}V^*_{cd}V_{us}(T''-A'')$
     & $1.72 \pm 0.19$  & $1.59 \pm 0.15$ \\
     & ${K}^{+} \eta$ & $V^*_{cd}V_{us}({1\over\sqrt{2}}(T''+A'')\cos\phi-A''\sin\phi)$
     &   & $0.98 \pm 0.04$ \\
     & ${K}^{+} \eta'$ &  $V^*_{cd}V_{us}({1\over\sqrt{2}}(T''+A'')\sin\phi+A''\cos\phi)$
     &   & $0.91 \pm 0.17$ \\
\hline
$D_s^+$ & ${K}^{0} K^+$ & $V^*_{cd}V_{us}(T''+C'' )$
     &  & $0.38 \pm 0.04$ \\
\end{tabular}
\end{ruledtabular}
\end{table}

A global fit to the data of singly Cabibbo-suppressed decay modes shows that the value of $\chi^2$ is very large, of order 87 per degree of freedom.  One solution we find is
\be \label{eq:PPall1}
&& T'=1.14, \qquad\qquad\qquad C'=2.36\,e^{i222^\circ}, \non \\
&&  E'=1.85\,e^{-i52^\circ},
\qquad\quad ~A'=2.51\,e^{i100^\circ}.
\en
This deviates substantially from the unprimed solution given in Eq.~(\ref{eq:PP1}).  The main processes contributing to large $\chi^2$ are $D^0\to K^+K^-$ ($\Delta\chi^2 = 462$), $D^0\to \pi^+\pi^-$ ($\Delta\chi^2 = 248$), $D^0\to \pi^0\pi^0$ ($\Delta\chi^2 = 119$), and $D^+ \to \pi^+ \eta$ ($\Delta\chi^2 = 96$).  This is an indication of large SU(3) breaking effects in these decay modes.  However, if we perform a fit restricted to the measured branching fractions of $D^0\to \pi^0\eta,~\pi^0\eta',~\eta\eta,~\eta\eta'$, the extracted amplitudes are (in units of $10^{-6}$ GeV and for $\phi = 40.4^\circ$)
\be
\label{eq:limited}
C'=2.88^{+0.14}_{-0.15}, \qquad E'= (1.37^{+0.15}_{-0.17})\,
e^{-i(79 \pm 5)^\circ},
\en
with $\chi^2=0.29$, where we have assumed $C$ to be real and the phase on $E$ is relative to $C$.  Note that there is a two-fold ambiguity in the relative strong phase, and we have picked the negative one that is consistent with Eq.~(\ref{eq:PP1}).  These values agree with the unprimed values in Eq.~(\ref{eq:PP1}) within 1-$\sigma$.  If the data of $D^0\to \pi^0\pi^0$ is also included in the above fit, it will dominate the $\chi^2$ value which now reads $19.6$.  A global fit to both CF and SCS modes yields $\chi^2=52$ per degree of freedom.

Assuming that the primed amplitudes are identical to unprimed ones, {\it i.e.}, $T'=T,~C'=C$, etc., the predicted branching fractions of SCS $D\to PP$ decays are shown in the last column of Table~\ref{tab:CSPP}. It is clear that the predicted rates for $\pi^+\pi^-$ and $\pi^0\pi^0$ are too large, while those for $K^+K^-$, $\pi^+\pi^0$, $\pi^+\eta$, $\pi^+\eta'$, $K^+\eta$ and $K^+\eta'$ are too small compared to experiments. The decay $D^0\to K^0\bar K^0$ is prohibited by SU(3) symmetry, but the measured rate is comparable to that of $D^0\to \pi^0\pi^0$.  In the next section we shall study flavor SU(3) symmetry violation in SCS decays using the factorization approach and investigate its possible connection to final-state interactions.  Setting the double-primed topological amplitudes as unprimed ones, the branching fractions of DCS processes are displayed in Table~\ref{tab:DCSPP}.

A few remarks on our assumption of ignoring the $SE$ and $SA$ amplitudes are in place here. The former (latter) diagram appears only in the singly Cabibbo-suppressed decay modes $D^0 \to \pi^0 \eta^{(\prime)}$ and $D^0 \to \eta \eta^{(\prime)}$ ($D^+ \to \pi^+ \eta^{(\prime)}$ and $D_s^+ \to K^+ \eta^{(\prime)}$).  Ref.~\cite{RosnerPP09} includes them in their analysis.  Two solutions have been found for the magnitude of $SE$.  One is relatively large when compared to the other amplitudes ($T,C,E$, and $A$) and thus argued unlikely.  The other is close to zero, and explains our good fit of $C'$ and $E'$ given in Eq.~(\ref{eq:limited}) even without including $SE$.  As to the magnitude of $SA$, Ref.~\cite{RosnerPP09} finds only one large solution that is again unlikely in comparison with the other amplitudes.  Moreover, our predictions for the branching fractions of the $D^+ \to \pi^+ \eta^{(\prime)}$ and $D_s^+ \to K \eta^{(\prime)}$ decays in Table~\ref{tab:CSPP}, which are based on the fit to Cabibbo-favored modes without including the $SA$ amplitude, are in better agreement with the data than the predictions given in Table~VI of Ref.~\cite{RosnerPP09} where the large $SA$ solution is used.  Therefore, the current data do not suggest significant $SE$ and $SA$ contributions.

The quantities $R(D^0)$, $R(D^+)$ and $R(D_s^+)$ defined by
\be
R(D^0) &=& {\Gamma(D^0\to K_S\pi^0)-\Gamma(D^0\to K_L\pi^0)\over \Gamma(D^0\to K_S\pi^0)+\Gamma(D^0\to K_L\pi^0)}\,,  \non \\
R(D^+) &=& {\Gamma(D^+\to K_S\pi^+)-\Gamma(D^+\to K_L\pi^+)\over \Gamma(D^+\to K_S\pi^+)+\Gamma(D^+\to K_L\pi^+)}\,,  \non \\
R(D^+_s) &=& {\Gamma(D^+_s\to K_S K^+)-\Gamma(D^+_s\to K_L K^+)\over \Gamma(D^+_s\to K_S K^+)+\Gamma(D^+_s\to K_L K^+)}
\en
measure the asymmetries between $K_S$ and $K_L$ production in these decays due to the interference between CF and DCS amplitudes. It is straightforward to show that \footnote{The formula $R(D^0)=2\tan^2\theta_C$ was originally derived under the phase convention such that $K_S={1\over\sqrt{2}}(K^0-\bar K^0)$ and $K_L={1\over\sqrt{2}}(K^0+\bar K^0)$ \cite{Bigi}. Of course, physics should be independent of the choice of phase convention. If we choose the convention $CP|K^0(\vec{p})\ra=|\bar K^0(-\vec{p})\ra$, we will have $K_S={1\over\sqrt{2}}(K^0+\bar K^0)$ and $K_L={1\over\sqrt{2}}(K^0-\bar K^0)$ in the absence of {\it CP} violation. Since the axial vector current $A_\mu=\bar s\gamma_\mu(1-\gamma_5)d$ transforms as $CP(\bar s\gamma^\mu(1-\gamma_5)d)(CP)^\dagger=-\bar d\gamma_\mu(1-\gamma_5)s$ under the {\it CP} transformation, it is clear that the decay constants of $K^0$ and $\bar K^0$ are opposite in sign. Consequently, $\A(D^0\to K^0\pi^0)=-(V_{cd}^*V_{us})/(V_{cs}^*V_{ud})\A(D^0\to \bar K^0\pi^0)=\tan^2\theta_C \A(D^0\to \bar K^0\pi^0)$. It follows that $D^0\to \bar K^0\pi^0$ and $D^0\to K^0\pi^0$ contribute constructively to $D^0\to K_S\pi^0$ and destructively to $D^0\to K_L\pi^0$.
}
\be
R(D^0) &=& 2\tan^2\theta_C=  0.107\,, \non \\
R(D^+) &=& 2\tan^2\theta_C\,{\rm Re}\left({C''+A''\over C+T}\right)
=-0.019\pm0.016 \,, \non \\
R(D_s^+) &=& 2\tan^2\theta_C\,{\rm Re}\left({C''+T''\over C+A}\right)
=-0.008\pm0.007\,.
\en
Note that $R(D^+)$ is predicted to be $0.035\sim 0.044$ in \cite{Gao} and $-0.005\pm0.013$ in \cite{RosnerPP09}, while $R(D_s^+)$ is $-0.0022\pm0.0087$ in \cite{RosnerPP09}. Our results for $R(D^+)$ and $R(D^+_s)$ differ in central values from those in \cite{RosnerPP09} since the latter were obtained using the topological amplitudes extracted for $\phi=43^\circ$.  The CLEO measurements are \cite{CLEO:RD0}
\be
R(D^0) = 0.108\pm0.025\pm0.024, \qquad R(D^+)=0.022\pm0.016\pm0.018.
\en
Evidently, the calculated $R(D^0)$ agrees well with experiment, while there is no evidence for a significant asymmetry in $D^+\to K_{S,L}\pi^+$ decays.

\subsection{$D\to VP$}

\begin{table}[tp!]
\caption{Branching fractions and invariant amplitudes for Cabibbo-favored decays of charmed mesons to one pseudoscalar and one vector meson.  Data are taken from \cite{PDG}.  The last two columns show the fitted branching fractions obtained from solutions (A, A1) and (S, S1), respectively, for $\phi = 40.4^\circ$.  Due to the lack of information on $A_P$ and $A_V$, no prediction is attempted for $D_s^+$ decays except $D_s^+\to\pi^+\phi$.
\label{tab:CFPV}}
\vspace{6pt}
\begin{ruledtabular}
\begin{tabular}{l l l c c c c}
Meson & Mode & Representation
 & ${\cal B}_{\rm exp}$ & ${\cal B}_{\rm fit}$ (A,A1) & ${\cal B}_{\rm fit}$ (S,S1) \\
 & & & $(\%)$ & $(\%)$ & $(\%)$ \\
\hline
$D^0$ & $K^{*-} \pi^+$ & $V^*_{cs}V_{ud}(T_V + E_P)$
     & $5.91 \pm 0.39$ & $5.91 \pm 0.70$ & $5.91 \pm 0.66$   \\
 & $K^- \rho^+$ & $V^*_{cs}V_{ud}(T_P + E_V)$
     & $10.8 \pm 0.7$ & $10.8 \pm 2.2$ & $10.7 \pm 2.3$ \\
 & $\ol{K}^{*0} \pi^0$ & $\frac{1}{\sqrt{2}}V^*_{cs}V_{ud}(C_P - E_P)$
     & $2.82 \pm 0.35$ & $2.82 \pm 0.34$ & $2.82 \pm 0.28$ \\
 & $\ol{K}^0 \rho^0$ & $\frac{1}{\sqrt{2}}V^*_{cs}V_{ud}(C_V - E_V)$
     & $1.54\pm0.12$ & $1.54 \pm 1.15$ & $1.55 \pm 0.34$ \\
 & $\ol{K}^{*0} \eta$ & $V^*_{cs}V_{ud}({1\over \sqrt{2}}(C_P + E_P)\cos\phi - E_V\sin\phi\,)$
     & $0.96 \pm 0.30 $ & $0.96 \pm 0.32$ & $1.12 \pm 0.26$ \\
 & $\ol{K}^{*0} \eta\,'$ & $V^*_{cs}V_{ud}({1\over \sqrt{2}}(C_P + E_P)\sin\phi - E_V\cos\phi\,)$
     & $< 0.11$ &  $0.012 \pm 0.003$ & $0.020 \pm 0.003$ \\
 & $\ol{K}^0 \omega$ & $\frac{1}{\sqrt{2}}V^*_{cs}V_{ud}(C_V + E_V)$
     & $2.26 \pm 0.40$ & $2.26 \pm 1.38$ & $2.34 \pm 0.41$ \\
 & $\ol{K}^0 \phi$ & $V^*_{cs}V_{ud}E_P$
     & $0.868 \pm 0.060$ &  $0.868 \pm 0.139$ & $0.868 \pm 0.110$ \\
\hline
$D^+$ & $\ol{K}^{*0} \pi^+$ & $V^*_{cs}V_{ud}(T_V + C_P)$
     & $1.83 \pm 0.14$ &  $1.83 \pm 0.49$ & $1.83 \pm 0.46$ \\
 & $\ol{K}^0 \rho^+$ & $V^*_{cs}V_{ud}(T_P + C_V)$
     & $9.2 \pm 2.0$ & $9.2 \pm 6.7$ & $9.7 \pm 5.2$ \\
\hline
$D_s^+$ & $\ol{K}^{*0} K^+$ & $V^*_{cs}V_{ud}(C_P + A_V)$
     & $3.91 \pm 0.23$ \footnotemark[1] &   \\
 & $\ol{K}^0 K^{*+}$ & $V^*_{cs}V_{ud}(C_V + A_P)$
     & $5.3 \pm 1.2$ &   \\
& $\rho^+ \pi^0$ & $\frac{1}{\sqrt{2}}V^*_{cs}V_{ud}(A_P - A_V)$
     & --- &  \\
 & $\rho^+ \eta$ & $V^*_{cs}V_{ud}({1\over\sqrt{2}}( A_P + A_V)\cos\phi-T_P\sin\phi)$
     & $8.9 \pm 0.8$ \footnotemark[2] &  \\
 & $\rho^+ \eta\,'$ & $V^*_{cs}V_{ud}({1\over\sqrt{2}}( A_P + A_V)\sin\phi+T_P\cos\phi)$
     & $12.2 \pm 2.0$  &  \\
 & $\pi^+ \rho^0$ & $\frac{1}{\sqrt{2}}V^*_{cs}V_{ud}(A_V - A_P)$
     & --- & \\
 & $\pi^+ \omega$ & $\frac{1}{\sqrt{2}}V^*_{cs}V_{ud}(A_V + A_P)$
     & $0.21 \pm 0.09$ \footnotemark[3] & \\
 & $\pi^+ \phi$ & $V^*_{cs}V_{ud}T_V$
     & $4.38 \pm 0.35$ & $4.38 \pm 0.35$ & $4.38 \pm 0.35$   \\
\end{tabular}
\end{ruledtabular}
\footnotetext[1]{From the fit fraction $\Gamma(D_s^+\to\bar K^{*0}K^+)/\Gamma(D_s^+\to K^+K^-\pi^+)=(47.4\pm1.5)\%$ \cite{CLEO:DsKKpi} combined with the PDG value \cite{PDG}.}
\footnotetext[2]{From \cite{CLEOrhoeta}.}
\footnotetext[3]{From \cite{CLEOomega}.}
\end{table}
\renewcommand{\arraystretch}{0.85}
\begin{table}
\caption{Same as Table~\ref{tab:CFPV} except for singly-Cabibbo-suppressed decays of charmed mesons.  Due to the lack of information on $A_P$ and $A_V$, no prediction is attempted for $D_s^+$ decays except $D^+\to\pi^+\phi$.
  \label{tab:CSPV}}
\begin{ruledtabular}
\begin{tabular}{l l l c c c c}
Meson & Mode & Representation & ${\cal B}_{\rm exp}$
 & ${\cal B}_{\rm theory}$ (A,A1) & ${\cal B}_{\rm theory}$ (S,S1) \\
 & & & $(\times 10^{-3})$ & $(\times 10^{-3})$ & $(\times 10^{-3})$ \\
\hline
$D^0$
  & $\pi^+ \rho^-$ & $V^*_{cd}V_{ud}(T_V'+E_P')$
     & $4.97 \pm 0.23$ & $3.92 \pm 0.46$ & $5.18 \pm 0.58$ \\
  & $\pi^- \rho^+$ & $V^*_{cd}V_{ud}(T_P'+E_V')$
     & $9.8 \pm 0.4$ & $8.34 \pm 1.69$ & $8.27 \pm 1.79$ \\
  & $\pi^0 \rho^0$ & $\frac12V^*_{cd}V_{ud}(C_P'+C_V'-E_P'-E_V')$
     & $3.73 \pm 0.22$ & $2.96 \pm 0.98$ & $3.34 \pm 0.33$ \\
  & $K^+ K^{*-}$ & $V^*_{cs}V_{us}(T'_V+E'_P)$
     & $1.53 \pm 0.15$ & $1.99 \pm 0.24$ & $1.99 \pm 0.22$ \\
  & $K^- K^{*+}$ & $V^*_{cs}V_{us}(T'_P+E'_V)$
     & $4.41 \pm 0.21$ & $4.25 \pm 0.86$ & $3.18 \pm 0.69$ \\
  & $K^0 \ol{K}^{*0}$ & $V^*_{cs}V_{us}E'_P+V^*_{cd}V_{ud}E'_V$
     & $<1.8$ & $0.29 \pm 0.22$ & $0.05 \pm 0.06$ \\
  & $\ol{K}^0 K^{*0}$ & $V^*_{cs}V_{us}E'_V+V^*_{cd}V_{ud}E'_P$
     & $<0.9$ & $0.29 \pm 0.22$ & $0.05 \pm 0.06$ \\
  & $\pi^0 \omega$ & $-\frac{1}{2}V^*_{cd}V_{ud}(C'_V-C'_P+E'_P+E'_V)$
     & $<0.26$ & $0.10 \pm 0.18$ & $1.01 \pm 0.18$ \\
  & $\pi^0 \phi$ & $\frac{1}{\sqrt{2}}V^*_{cs}V_{us}C'_P$
     & $1.24 \pm 0.12$ & $1.22 \pm 0.08$ & $1.11 \pm 0.05$ \\
  & $\eta \omega$ & $V^*_{cd}V_{ud}{1\over 2}(C'_V+C'_P+E'_V+E'_P)\cos\phi$
     & $2.21\pm0.23$ \footnotemark[1] & $3.08 \pm 1.42$ & $3.94 \pm 0.61$ \\
  & & $\qquad-V^*_{cs}V_{us}{1\over\sqrt{2}}C'_V\sin\phi$ &&&& \\
  & $\eta\,' \omega$ & $V^*_{cd}V_{ud}{1\over 2}(C'_V+C'_P+E'_V+E'_P)\sin\phi$
     &--- & $0.07 \pm 0.02$ & $0.15 \pm 0.01$ \\
   & & $\qquad+V^*_{cs}V_{us}{1\over\sqrt{2}}C'_V\cos\phi$ &&&& \\
  & $\eta \phi$ & $V^*_{cs}V_{us}({1\over\sqrt{2}}C'_P\cos\phi-(E'_V+E'_P)\sin\phi)$
     & $0.14\pm0.05$ & $0.31 \pm 0.10$ & $0.41 \pm 0.08$ \\
  & $\eta \rho^0$ & $V^*_{cd}V_{ud}{1\over 2}(C'_V-C'_P-E'_V-E'_P)\cos\phi$
     & --- &  $1.11 \pm 0.86$ & $1.17 \pm 0.34$ \\
    & & $\qquad -V^*_{cs}V_{us}{1\over\sqrt{2}}C'_V\sin\phi$ &&&& \\
  & $\eta\,' \rho^0$ & $V^*_{cd}V_{ud}{1\over 2}(C'_V-C'_P-E'_V-E'_P)\sin\phi$
     & --- & $0.14 \pm 0.02$ & $0.26 \pm 0.02$ \\
      & & $\qquad +V^*_{cs}V_{us}{1\over\sqrt{2}}C'_V\cos\phi$ &&&& \\
\hline
$D^+$
  & $\pi^+ \rho^0$ & $\frac{1}{\sqrt{2}}V^*_{cd}V_{ud}(T'_V+C'_P-A'_P+A'_V)$
     & $0.82 \pm 0.15$ &  \\
  & $\pi^0 \rho^+$ & $\frac{1}{\sqrt{2}}V^*_{cd}V_{ud}(T'_P+C'_V+A'_P-A'_V)$
     &--- &  \\
  & $\pi^+ \omega$ & $\frac{1}{\sqrt{2}}V^*_{cd}V_{ud}(T'_V+C'_P+A'_P+A'_V)$
     &$2.1\pm0.9$ \footnotemark[2] &  \\
  & $\pi^+ \phi$ & $V^*_{cs}V_{us}C'_P$
     &$6.2 \pm 0.7$  & $6.21 \pm 0.43$ & $5.68 \pm 0.28$ \\
  & $\eta \rho^+$ & $V^*_{cd}V_{ud}{1\over\sqrt{2}}(T'_P+C'_V+A'_V+A'_P)\cos\phi$
     & $<7$ &  \\
  &&$\qquad-V^*_{cs}V_{us}C'_V\sin\phi$&&&& \\
  & $\eta\,' \rho^+$ & $V^*_{cd}V_{ud}{1\over\sqrt{2}}(T'_P+C'_V+A'_V+A'_P)\sin\phi$
     & $<5$ &  \\
 &&$\qquad+V^*_{cs}V_{us}C'_V\cos\phi$&&&& \\
  & $K^+ \ol{K}^{*0}$ & $V^*_{cd}V_{ud}A'_V+V^*_{cs}V_{us}T'_V$
     & $4.4 \pm 0.5$ &  \\
  & $\ol{K}^0 K^{*+}$ & $V^*_{cd}V_{ud}A'_P+V^*_{cs}V_{us}T'_P$
     & $31.8 \pm 13.8$ &  \\
\hline
$D_s^+$
  & $\pi^+ K^{*0}$ & $V^*_{cd}V_{ud}T'_V+V^*_{cs}V_{us}A'_V$
     & $2.25 \pm 0.39$ &  \\
  & $\pi^0 K^{*+}$ & $\frac{1}{\sqrt{2}}(V^*_{cd}V_{ud}C'_V-V^*_{cs}V_{us}A'_V)$
     & --- &  \\
  & $K^+ \rho^0$ & $\frac{1}{\sqrt{2}}(V^*_{cd}V_{ud}C'_P-V^*_{cs}V_{us}A'_P)$
     & $2.7 \pm 0.5$ &  \\
  & $K^0 \rho^+$ & $V^*_{cd}V_{ud}T'_P+V^*_{cs}V_{us}A'_P$
     & --- &  \\
  & $\eta K^{*+}$ & ${1\over\sqrt{2}}(V^*_{cd}V_{ud}C'_V+V^*_{cs}V_{us}A'_V)\cos\phi$
     & --- &  \\
  &&$\qquad-V^*_{cs}V_{us}(T'_P+C'_V+A'_P)\sin\phi$&&&& \\
  & $\eta\,' K^{*+}$
  & ${1\over\sqrt{2}}(V^*_{cd}V_{ud}C'_V+V^*_{cs}V_{us}A'_V)\sin\phi$
     & --- &  \\
   &&$\qquad+V^*_{cs}V_{us}(T'_P+C'_V+A'_P)\cos\phi$&  &&& \\
  & $K^+ \omega$ & $\frac{1}{\sqrt{2}}\left(V^*_{cd}V_{ud}C'_P
  	+V^*_{cs}V_{us}A'_P\right)$
     & $<2.4$ \footnotemark[2] &  \\
  & $K^+ \phi$ & $V^*_{cs}V_{us}(T'_V+C'_P+A'_V)$
     & $<0.57$  &  \\
\end{tabular}
\footnotetext[1]{Data from \cite{Kass}.}
\footnotetext[2]{Data from \cite{CLEOomega}.}
\end{ruledtabular}
\end{table}
\begin{table}
\caption{Same as Table~\ref{tab:CFPV} except for doubly-Cabibbo-suppressed decays of charmed mesons.
  \label{tab:DCSPV}}
\begin{ruledtabular}
\begin{tabular}{l l l c c c c}
Meson & Mode & Representation & ${\cal B}_{\rm exp}$
& ${\cal B}_{\rm theory}$ (A,A1) & ${\cal B}_{\rm theory}$ (S,S1) \\
 & & & $(\times 10^{-4})$ & $(\times 10^{-4})$ & $(\times 10^{-4})$ \\
\hline
$D^0$
& $K^{*+}\,\pi^-$ & $V^*_{cd}V_{us}(T''_P+E''_V)$ & $3.0^{+3.9}_{-1.2}$
& $3.59 \pm 0.72$ & $2.69 \pm 0.58$ \\
& $K^{*0}\,\pi^0$ & $\frac{1}{\sqrt{2}}V^*_{cd}V_{us}\left( C''_P-E''_V \right)$ &
& $0.54 \pm 0.18$ & $0.74 \pm 0.17$ \\
& $\phi\,K^0$ & $V^*_{cd}V_{us}E''_V$ &
& $0.06 \pm 0.05$ & $0.15 \pm 0.06$ \\
& $\rho^-\,K^+$ & $V^*_{cd}V_{us}(T''_V+E''_P)$ &
& $1.45 \pm 0.17$ & $1.91 \pm 0.21$ \\
& $\rho^0\,K^0$ & $\frac{1}{\sqrt{2}}V^*_{cd}V_{us}(C''_V-E''_P)$ &
& $0.91 \pm 0.51$ & $0.63 \pm 0.19$ \\
& $\omega\,K^0$ & $\frac{1}{\sqrt{2}}V^*_{cd}V_{us}(C''_V+E''_P)$ &
& $0.58 \pm 0.40$ & $0.85 \pm 0.21$ \\
& $K^{*0}\,\eta$ & $V^*_{cd}V_{us}({1\over\sqrt{2}}(C''_P+E''_V)\cos\phi-E''_P\sin\phi)$ &
& $0.33 \pm 0.08$ & $0.28 \pm 0.05$ \\
& $K^{*0}\,\eta^{\prime}$ & $V^*_{cd}V_{us}({1\over\sqrt{2}}(C''_P+E''_V)\sin\phi+E''_P\cos\phi)$ &
& $0.0040 \pm 0.0006$ & $0.0061 \pm 0.0004$ \\
\hline
$D^+$
& $K^{*0}\,\pi^+$ & $V^*_{cd}V_{us}(C''_P+A''_V)$ & $4.35 \pm 0.90$ &  & $$ \\
& $K^{*+}\,\pi^0$ & $\frac{1}{\sqrt{2}}V^*_{cd}V_{us}\left( T''_P-A''_V \right)$ & &  & \\
& $\phi\,K^+$ & $V^*_{cd}V_{us}A''_V$ & &  & \\
& $\rho^+\,K^0$ & $V^*_{cd}V_{us}(C''_V+A''_P)$ & &  & \\
& $\rho^0\,K^+$ & $\frac{1}{\sqrt{2}}V^*_{cd}V_{us}(T''_V-A''_P)$
& $2.4 \pm 0.6$ & & $$ \\
& $\omega\,K^+$ & $\frac{1}{\sqrt{2}}V^*_{cd}V_{us}(T''_V+A''_P)$ &  & & \\
& $K^{*+}\,\eta$ & $V^*_{cd}V_{us}({1\over\sqrt{2}}(T''_P+A''_V)\cos\phi-A''_P\sin\phi)$  &  &  & \\
& $K^{*+}\,\eta^{\prime}$ & $V^*_{cd}V_{us}({1\over\sqrt{2}}(T''_P+A''_V)\sin\phi+A''_P\cos\phi)$
& &  & \\
\hline
$D^+_s$
& $K^{*+}\,K^0$ & $V^*_{cd}V_{us}(T''_P+C''_V)$ &
& $1.17 \pm 0.86$ & $1.03 \pm 0.55$  \\
& $K^{*0}\,K^+$ & $V^*_{cd}V_{us}(T''_V+C''_P)$ &
& $0.20 \pm 0.05$ & $0.22 \pm 0.06$  \\
\end{tabular}
\end{ruledtabular}
\end{table}

The topological amplitude expressions for CF, SCS and DCS $D\to VP$ decays are listed in Tables \ref{tab:CFPV} to \ref{tab:DCSPV}, respectively.  For reduced amplitudes $T$ and $C$ in $D\to VP$ decays, the subscript $P$ ($V$) implies a pseudoscalar (vector) meson which contains the spectator quark of the charmed meson.  For $E$ and $A$ amplitudes with the final state $q_1\bar q_2$, the subscript $P$ ($V$) denotes a pseudoscalar (vector) meson which contains the antiquark $\bar q_2$.

The invariant amplitude is related to the partial width via
 \be \label{eq:VP1}
 \Gamma(D\to VP)={p_c\over 8\pi m_D^2}\sum_{pol.}|\A|^2
 \en
by summing over the the polarization states of the vector meson, or through the relation
 \be \label{eq:VP2}
 \Gamma(D\to VP)={p^3_c\over 8\pi m_D^2}|\tilde \A|^2,
 \en
by taking the polarization vector out of the amplitude, where $\A= (m_V/m_D)\tilde \A\, (\varepsilon\cdot p_D)$. The first approach has been used in \cite{Chiang04} for the extraction of topological amplitudes in $B\to VP$ decays, while the second approach was employed in \cite{Rosner99,Chiang03,RosnerVP} for $D\to VP$ . To illustrate the relations between Eqs.~(\ref{eq:VP1}) and (\ref{eq:VP2}), we consider the channel $D_s^+\to\pi^+\phi$ as an example. Its amplitude is given by $\A=V_{cs}^*V_{ud}T_V$ with
\be
T_V=2 f_\pi m_{\phi}A_0^{D_s\phi}(m_\pi^2)(\varepsilon\cdot p_D).
\en
evaluated in the factorization approach. After summing over the polarizations of the vector meson, the above expression can be simplified by replacing $m_V\,\varepsilon\cdot p_D$ with $m_D p_c$. Hence, we obtain Eq.~(\ref{eq:VP2}) with
\be \label{eq:A&S}
\tilde T_V=2 f_\pi m_D A_0^{D_s\phi}(m_\pi^2)={m_D\over m_V}\,{T_V\over \varepsilon\cdot p_D}.
\en

From the decays $D_s^+\to\pi^+\phi$, $D^0\to\bar K^0\phi$ and the three CF $D\to
\bar K^*\pi$ channels, two solutions for the magnitudes of and relative phases between $T_V$, $C_P$ and $E_P$ are found and shown in Table~\ref{tab:TV} for two different ways of extracting topological amplitudes using (\ref{eq:VP1}) and
(\ref{eq:VP2}).  It is noted that these results have no dependence on the $\eta$-$\eta'$ mixing angle.  Solutions with a smaller $C_P$, {\it i.e.} A' and S', are ruled out by the measurements of SCS decays $D^0\to \pi^0\phi$ and $D^+\to\pi^+\phi$ since they will lead to predictions too small by a factor of 3 when confronted with experiment, \footnote{There are two different values of $\B(D^0\to\pi^0\phi)$ quoted by the Particle Data Group \cite{PDG}: $(1.24\pm0.12)\times 10^{-3}$ obtained by BaBar \cite{BaBar:piphi} and CLEO \cite{CLEO:piphi} from the Dalitz-plot analysis with interference, and $(0.76\pm0.05)\times 10^{-3}$ obtained by Belle by measuring the background for the radiative decay $D^0\to\phi\gamma$ \cite{Belle:piphi}. In this work, we shall use the former.}
while the data can be nicely explained with the larger $C_P$ (see Table \ref{tab:CSPV}), as noticed in \cite{Wu04,RosnerVP}.  Although the fact that the magnitude of color-suppressed $C_P$ is larger than the color-allowed $T_V$ seems to be in contradiction with the na{\"i}ve expectation, we shall see in the next section that the corresponding effective parameter $a_1$ is still larger than $a_2$, as anticipated from the short-distance approach.

Using the solutions of $T_V$, $C_P$ and $E_P$ as inputs, the other amplitudes $T_P$, $C_V$ and $E_V$ can be obtained by considering the decay modes $D^0\to\bar K^0\omega$, $D^0\to \bar K^{*0}\eta$, $D^0\to K^-\rho^+,~\bar K^0\rho^0$ and $D^+\to \bar K^0\rho^+$.  Here we have assumed that $T_P$ and $T_V$ are relatively real.  We obtain 6 solutions A1$-$A6 for topological amplitudes extracted from (\ref{eq:VP2}) and 5 solutions S1$-$S5 when amplitudes are extracted from (\ref{eq:VP1}), as given in Table~\ref{tab:TP}.

Several remarks are in order: (i) All the solutions are exact with $\chi^2=0$, except for S1 where the fit $\chi^2$ value is 0.42.  This is reflected in some of the predicted branching fractions in Table~\ref{tab:CFPV}.    (ii) Solutions displayed in Table \ref{tab:TP} are obtained using the $\eta-\eta'$ mixing angle $\phi=40.4^\circ$.  If the angle $\phi=35.26^\circ$ is employed, we obtain
\be
&& (A1)~~T_P=7.85^{+0.54}_{-2.27},
\quad C_V=(3.63^{+0.71}_{-1.21})\,e^{i(172^{+16}_{-14})^\circ},
\quad E_V=(2.51^{+1.19}_{-1.18})e^{-i(110^{+16}_{-23})^\circ} ~, \non \\
&& (S1)~~T_P = 3.12^{+0.29}_{-0.39},
\quad C_V=(1.25^{+0.32}_{-0.39})\,e^{i(177^{+13}_{-11})^\circ},
\quad E_V = (1.42^{+0.29}_{-0.34})e^{-i(107^{+10}_{-11})^\circ} ~.
\en
where the amplitudes in the A1 solution are quoted in units of $10^{-6}$ and those in the S1 solution in units of $10^{-6} (\varepsilon \cdot p_D)$.  By comparing them with solutions A1 and S1 in Table~\ref{tab:TP}, we see that a small decrease of the $\eta\!-\!\eta'$ mixing angle will reduce $T_P$ and $C_V$ slightly and enhance the magnitude of $E_V$.   (iii) Using $\phi=35.26^\circ$ and removing the CKM angles $V_{cd}^*V_{ud}$ from the amplitudes, we are able to reproduce solutions A1$-$A6 in Table III of \cite{RosnerVP}. \footnote{It seems to us that there is a sign typo in solution A6 of \cite{RosnerVP}: The amplitude $T_P$ there should be negative.}  (iv) The relation $E_V=-E_P$ used in \cite{Chiang03} is not borne out in this analysis. Instead, the angle between $E_V$ and $E_P$ turns out to be small.  (v) To see the relative phase $\delta_{T_VT_P}$ between $T_P$ and $T_V$, we fit to the data of the CF $D^0$, $D^+$ decays and the channel $D_s^+\to \pi^+\phi$ and find that $\delta_{T_VT_P}$ is of order $(16 \sim 18)^\circ$ when using Eq.~(\ref{eq:VP1}) and $(-1 \sim 1)^\circ$ when using Eq.~(\ref{eq:VP2}).  Therefore, the assumption of relatively real $T_P$ and $T_V$ amplitudes is justified.   (vi) A careful comparison between the two sets of solutions (A1$-$A6 and S1$-$S5) in Table~\ref{tab:TP} shows that there is no solution in the latter set that corresponds to solution A2. Ref.~\cite{RosnerVP}
claims that solution A2 is an alternative solution to A1 as it
produces roughly the same $\chi^2$ as A1 when they are applied to the SCS decays.  However, we find that the $D^0 \to \phi \eta$ mode alone contributes $\Delta\chi^2 = 285$ in solution A2, sufficient to disregard this solution.

\begin{table}[t]
\caption{Solutions for $T_V$, $C_P$ and $E_P$. The top two solutions in units of $10^{-6}$ are obtained using Eq.~(\ref{eq:VP2}), while the bottom two solutions in units of $10^{-6}(\varepsilon \cdot p_D)$ are extracted using Eq.~(\ref{eq:VP1}).} \label{tab:TV}
\begin{ruledtabular}
\begin{tabular}{ l c c c }
& $T_V$ & $C_P$ & $E_P$ \\
 \hline
A & $4.16^{+0.16}_{-0.17}$
& $(5.14^{+0.30}_{-0.33})e^{-i(162\pm3)^\circ}$
& $(3.09 \pm 0.11)e^{-i(93\pm5)^\circ}$ \\
A'& $4.16^{+0.16}_{-0.17}$
& $(3.00^{+0.35}_{-0.32})e^{-i(158^{+3}_{-4})^\circ}$
& $(3.09 \pm 0.11)e^{i(93\pm5)^\circ}$ \\
\hline
S & $2.15^{+0.08}_{-0.09}$
& $(2.68^{+0.14}_{-0.15})e^{-i(164\pm3)^\circ}$
& $(1.69 \pm 0.06)e^{-i(103\pm4)^\circ}$ \\
S'& $2.15^{+0.08}_{-0.09}$
& $(1.53^{+0.17}_{-0.15})e^{-i(162^{+3}_{-5})^\circ}$
& $(1.69 \pm 0.06)e^{i(103\pm4)^\circ}$ \\
\end{tabular}
\end{ruledtabular}
\end{table}

\begin{table}[t]
\caption{Solutions for $T_P$, $C_V$ and $E_V$ with inputs from solution A and S in Table~\ref{tab:TV}. Solutions A1$-$A6 in units of $10^{-6}$ are obtained using Eq.~(\ref{eq:VP2}), while solutions S1$-$S5 in units of $10^{-6}(\varepsilon \cdot p_D)$ are extracted using Eq.~(\ref{eq:VP1}).  Here we take $\phi = 40.4^\circ$.}
\label{tab:TP}
\begin{ruledtabular}
\begin{tabular}{l c c c }
 & $T_P$ & $C_V$ & $E_V$ \\
 \hline
A1 & $8.11^{+0.32}_{-0.43}$ &  $(4.15^{+0.34}_{-0.57})e^{i(164^{+36}_{-10})^\circ}$
& $(1.51^{+0.97}_{-0.69})e^{-i(124^{+57}_{-26})^\circ}$ \\
A2 & $6.16^{+0.55}_{-0.51}$ &  $(1.99^{+0.63}_{-0.60})e^{-i(165^{+9}_{-8})^\circ}$
& $(3.95^{+0.31}_{-0.41})e^{-i(89\pm6)^\circ}$ \\
A3 & $-6.59^{+0.79}_{-0.62}$ &  $(4.09^{+0.29}_{-0.39})e^{-i(39^{+12}_{-11})^\circ}$
& $(1.66^{+0.66}_{-0.59})e^{-i(113\pm16)^\circ}$ \\
A4 & $-6.39^{+1.05}_{-1.03}$ &  $(2.19^{+1.16}_{-1.07})e^{-i(12^{+14}_{-21})^\circ}$
& $(3.83^{+0.56}_{-1.03})e^{-i(89\pm10)^\circ}$ \\
A5 & $-3.33^{+0.34}_{-0.37}$ &  $(1.74^{+0.59}_{-0.76})e^{-i(111^{+15}_{-17})^\circ}$
& $(4.06^{+0.32}_{-0.38})e^{-i(185^{+11}_{-9})^\circ}$ \\
A6& $-3.46^{+0.39}_{-0.45}$ &  $(2.03^{+0.59}_{-0.79})e^{i(99^{+16}_{-13})^\circ}$
& $(3.93^{+0.38}_{-0.44})e^{-i(185\pm10)^\circ}$ \\
\hline
S1 & $3.14^{+0.31}_{-0.50}$ &  $(1.33^{+0.36}_{-0.51})e^{i(177^{+16}_{-13})^\circ}$
& $(1.31^{+0.40}_{-0.47})e^{-i(106^{+13}_{-16})^\circ}$ \\
S2 & $-2.20^{+0.96}_{-0.35}$ &  $(1.46^{+0.19}_{-0.30})e^{-i(53^{+14}_{-19})^\circ}$
& $(1.13^{+0.28}_{-0.26})e^{-i(131^{+15}_{-57})^\circ}$ \\
S3 & $-2.11^{+0.64}_{-0.37}$ &  $(0.46^{+0.43}_{-0.30})e^{-i(39^{+21}_{-65})^\circ}$
& $(1.79^{+0.12}_{-0.20})e^{-i(104^{+9}_{-12})^\circ}$ \\
S4 & $-1.41^{+0.17}_{-0.30}$ &  $(0.82^{+0.34}_{-0.35})e^{-i(102^{+26}_{-19})^\circ}$
& $(1.66^{+0.16}_{-0.78})e^{-i(177^{+22}_{-11})^\circ}$ \\
S5 & $-1.38^{+0.14}_{-0.15}$ &  $(0.75^{+0.24}_{-0.31})e^{i(109^{+18}_{-14})^\circ}$
& $(1.69^{+0.14}_{-0.15})e^{-i(177^{+13}_{-10})^\circ}$ \\
\end{tabular}
\end{ruledtabular}
\end{table}

Now we come to the remaining parameters $A_P$ and $A_V$ appearing in the CF $D_s^+$ decays. The non-observation of the decays $D_s^+\to \rho^0\pi^+$ and $\rho^+\pi^0$ gives a strong constraint on the $W$-annihilation amplitudes.  The mode $D_s^+\to \rho^0\pi^+$ was quoted by the Particle Data Group as ``not seen" \cite{PDG}.  While the ratio $\Gamma(D_s^+\to\rho^0\pi^+) / \Gamma(D_s^+\to\pi^+\pi^+\pi^-)$ was measured by E791 to be $(5.8\pm2.3\pm3.7)\%$ \cite{E791}, a most recent Dalitz plot analysis of $D_s^+\to\pi^+\pi^+\pi^-$ by BaBar yielded the fit fraction $(1.8\pm0.5\pm1.0)\%$ \cite{BaBarrhopi}. This mode was also not seen by the FOCUS Collaboration \cite{KLOE:Ds3pi}. Given the branching fraction $\B(D_s^+\to\pi^+\pi^+\pi^-)=(1.11\pm0.08)\%$ \cite{PDG}, the BaBar result leads to the upper limit $\B(D_s^+\to \rho^0\pi^+)<5\times 10^{-4}$.  If we ignore the decays $D_s^+\to \rho^0\pi^+$ and $\rho^+\pi^0$ for the moment, the annihilation amplitudes $A_P$ and $A_V$ can be determined from the four channels $\bar K^{*0}K^+$, $\bar K^0K^{*+}$, $\rho^+\eta$ and $\pi^+\omega$ in conjunction with the information of $C_P$ from solution A or S and $T_P,~C_V$ from solution A1 or S1. It turns out that although the above four $D_s^+$ modes are nicely fitted, the predicted $\B(D_s^+\to \rho^0\pi^+)$ of order $4\times 10^{-3}$ is too large. \footnote{In Ref.~\cite{RosnerVP}, the amplitudes $A_V$ and $A_P$ were determined from $D_s^+\to (\bar K^{*0}K^+,~\bar K^0K^{*+},~ \pi^+\omega)$.  It is not clear to us how the four unknown parameters (two magnitudes and two phases) can be extracted out of three data points.  Anyway, solutions A1 and A2 given in Table~IV of \cite{RosnerVP} for $A_P$ and $A_V$ lead to $\B(D_s^+\to \rho^0\pi^+)\sim 1.2\%$ and $2.2\%$ in their central values, respectively, which are obviously too large compared to experiment.}

From the measured rates of $D_s^+\to \pi^+\rho^0$ and $D_s^+\to\pi^+\omega$  we have
\be \label{eq:APAV1}
|A_V-A_P|<0.20\times 10^{-6}(\varepsilon\cdot p_D), \qquad |A_V+A_P|=(0.41\pm0.09)\times 10^{-6}(\varepsilon\cdot p_D)
\en
obtained using Eq.~(\ref{eq:VP1}) and
\be \label{eq:APAV2}
|A_V-A_P|<0.50\times 10^{-6}, \qquad |A_V+A_P|=(1.04\pm0.22)\times 10^{-6}
\en
using Eq.~(\ref{eq:VP2}).  Hence, the magnitudes of $A_P$ and $A_V$ are much smaller than that of $C_P$ and $C_V$.  In principle, the annihilation amplitudes can be determined when the above equation is combined with the amplitudes of $D_s^+\to \bar K^{*0}K^+$ and $\bar K^0K^{*+}$. However since $|C_P|>|C_V|\gg |A_P|, |A_V|$, it is not possible to have a nice fit to the data of $\bar K^{*0}K^+$, $\bar K^0K^{*+}$, $\pi^+\omega$ and $\pi^+\rho^0$ simultaneously. To see this, we set $A_V\approx A_P$ as a consequence of the non-observation of $D_s^+\to \rho^0\pi^+$. Under this relation, it is naively expected that $D_s^+\to \bar K^0K^{*+}$ has a rate larger than $D_s^+\to\bar K^{*0}K^+$ since $|C_P|>|C_V|$. Experimentally, it is the other way around: $\B(D_s^+\to \bar K^{*0}K^+)=(3.97\pm0.21)\%$ and $\B(D_s^+\to \bar K^{0}K^{*+})=(5.3\pm1.2)\%$. Since the former has been measured several times in the past decade (see Particle Data Group \cite{PDG} and the latest one \cite{CLEO:DsKKpi}) while the latter was measured two decades ago \cite{CLEO1989}, it is likely that the quoted experimental result for $D_s^+\to \bar K^{0}K^{*+}$ was overestimated. In short, a sensible determination of $A_P$ and $A_V$ cannot be made at present and we have to await more accurate measurement of $D_s^+\to \bar K^{0}K^{*+}$.

The cited experimental result for $D_s^+\to\rho^+\eta'$ is also problematic. Since $|T_P|\gg |A_V+A_P|$, we can neglect the annihilation contributions for the moment. It follows that if we use the solution with the largest $T_P$, the predicted branching ratios will be $\B(D_s^+\to\rho^+\eta)=(7.4 \pm 0.7)\%$ [$(7.1 \pm 1.8)\%$] and $\B(D_s^+\to\rho^+\eta')=(2.7 \pm 0.2)\%$ [$(2.6 \pm 0.7)\%$)] for solution A1 [S1]. While the predicted rate for the former is close to the recent CLEO measurement $\B(D_s^+\to\rho^+\eta)=(8.9\pm0.8)\%$, the latter is far below the quoted result $\B(D_s^+\to\rho^+\eta')=(12.2\pm2.0)\%$. It is very unlikely that the flavor-singlet contribution unique to the $\eta_0$ production can enhance its branching fraction from 3\% to 12\%.  This issue should be clarified by new measurement of $D_s^+\to\rho^+\eta'$.

Among the solutions A1$-$A6 and S1$-$S5, A1 and S1 are our preferred solutions for the following two reasons: (i) A global fit to the SCS decays of the $D^0$ meson (data of $D^+$ and $D_s^+$ are not included in the fit due to the lack of information on $A_P$ and $A_V$) indicates that A1 and S1 have the lowest values of $\chi^2$, 89 and 165, respectively, and (ii) $T_P$ has to be sufficiently large in order to possibly accommodate the data of $D_s^+\to \rho^+\eta$.

\section{Phenomenological Implications \label{sec:pheno}}

\subsection{Parameters $a_1$ and $a_2$}

In the diagrammatic approach, the topological amplitudes extracted using the $\chi^2$ fit method are not unique. Indeed, several other solutions are also allowed.  It is useful to consider the amplitudes for color-allowed and color-suppressed diagrams in the factorization approach in order to discriminate between different solutions. For CF $D\to \bar K\pi,~\bar K^*\pi,~\bar K\rho$ decays, the factorizable amplitudes read
 \be \label{eq:Ampfac1}
 T &=& {G_F\over
 \sqrt{2}}a_1(\ov K\pi)\,f_\pi(m_D^2-m_K^2)F_0^{DK}(m_\pi^2),
 \non \\
 C &=& {G_F\over
 \sqrt{2}}a_2(\ov K\pi)\,f_K(m_D^2-m_\pi^2)F_0^{D\pi}(m_K^2),
 \non \\
 T_V &=& {G_F\over
 \sqrt{2}}a_1(\ov K^*\pi)\,2f_\pi m_{K^*}A_0^{DK^*}(m_\pi^2)(\varepsilon\cdot p_D),
 \non \\
 C_P &=& {G_F\over
 \sqrt{2}}a_2(\ov K^*\pi)\,2f_{K^*}m_{K^*}F_1^{D\pi}(m_{K^*}^2)(\varepsilon\cdot p_D),
  \\
 T_P &=& {G_F\over
 \sqrt{2}}a_1(\ov K\rho)\,2f_\rho m_\rho F_1^{DK}(m_\rho^2)(\varepsilon\cdot p_D),
 \non \\
 C_V &=& {G_F\over
 \sqrt{2}}a_2(\ov K\rho)\,2f_K m_\rho
 A_0^{D\rho}(m_K^2)(\varepsilon\cdot p_D),\non
 \en
where we have followed \cite{BSW} for the definition of form factors. Factorization implies a universal, process-independent $a_1$ and $a_2$, for example, $a_2(\ov K\rho)=a_2(\ov K^*\pi)=a_2(\ov K\pi)$.

Under the na{\"i}ve factorization hypothesis, one has
 \be \label{nf}
a_1(\mu)=c_1(\mu)+{1\over N_c}c_2(\mu), \qquad \quad
a_2(\mu)=c_2(\mu)+{1\over N_c}c_1(\mu),
 \en
for decay amplitudes induced by current-current operators $O_{1,2}(\mu)$, where $c_{1,2}(\mu)$ are the corresponding Wilson coefficients and $N_c$ is the number of colors. However, it is well known that this naive factorization approach encounters two difficulties: (i) the coefficients $a_i$ given by Eq.  (\ref{nf}) are renormalization scale and $\gamma_5$-scheme dependent, and (ii) it fails to describe the color-suppressed decay modes due to the smallness of $a_2$. In particular, the ratio of $D^0\to \bar K^0\pi^0$ to $D^0\to K^-\pi^+$ is predicted to be of order $10^{-2}$, while experimentally it is close to 1/2.  Therefore, it is necessary to take into account nonfactorizable corrections:
\begin{eqnarray}  \label{a12}
a_1= c_1(\mu) + c_2(\mu) \left({1\over N_c} +\chi_1(\mu)\right)\,,
\qquad \quad a_2 = c_2(\mu) + c_1(\mu)\left({1\over N_c} +
\chi_2(\mu)\right)\,,
\end{eqnarray}
where nonfactorizable terms are characterized by the parameters $\chi_i$, which receive corrections including vertex corrections, hard spectator interactions involving the spectator quark of the heavy meson, and FSI effects from inelastic rescattering, resonance effects, etc.  The nonfactorizable terms $\chi_i(\mu)$ will compensate the scale and scheme dependence of Wilson coefficients to render $a_i$ physical.  In the so-called large-$N_c$ approach, a rule of discarding subleading $1/N_c$ terms is formulated \cite{Buras}. This amounts to having a universal nonfactorizable term $\chi_1=\chi_2=-1/N_c$ in Eq.~(\ref{a12}) and hence
 \be
 a_1\approx c_1(\bar m_c)= 1.274\,, \qquad\qquad a_2\approx c_2(\bar m_c)= -0.529
 \en
for $\Lambda_{\overline{\rm MS}}=215$ MeV and $\bar m_c(m_c)= 1.3$ GeV \cite{Buras96}.  This corresponds to a relative strong phase of $180^\circ$.  Empirically, this set of $a_1$ and $a_2$ gives a good description of the hadronic charm decays.  Hence, $a_1$ and $a_2$ in the large-$N_c$ approach can be considered as the theoretical benchmark values.

From the topological amplitudes obtained in Sec. II, we are ready to determine the effective Wilson coefficients $a_1$ and $a_2$ and their ratios. For the invariant amplitudes of $D\to VP$ determined from Eq.~(\ref{eq:VP1}), $a_{1,2}$ are extracted using the factorizable amplitudes given in Eq.~(\ref{eq:Ampfac1}). However, if the topological amplitudes of $D\to VP$ are extracted from Eq.~(\ref{eq:VP2}), we should use the following factorizable amplitudes
\be \label{eq:Ampfac2}
 \tilde T_V &=& {G_F\over
 \sqrt{2}}a_1(\ov K^*\pi)\,2f_\pi m_D A_0^{DK^*}(m_\pi^2),
 \non \\
 \tilde C_P &=& {G_F\over
 \sqrt{2}}a_2(\ov K^*\pi)\,2f_{K^*}m_{D}F_1^{D\pi}(m_{K^*}^2),
  \non \\
 \tilde T_P &=& {G_F\over
 \sqrt{2}}a_1(\ov K\rho)\,2f_\rho m_D F_1^{DK}(m_\rho^2),
 \non \\
 \tilde C_V &=& {G_F\over
 \sqrt{2}}a_2(\ov K\rho)\,2f_K m_D
 A_0^{D\rho}(m_K^2)
 \en
to determine $a_{1,2}$.  As noted in passing, the amplitudes $\tilde \A$ and $\A$ are related by Eq.~(\ref{eq:A&S}).

\begin{table}[t]
\caption{Form factors at $q^2=0$ and the shape parameter $\alpha$.}
\begin{ruledtabular} \label{tab:FF}
\begin{tabular}{l c c c c c c}
 & $F_0^{D\pi}$ & $F_0^{DK}$ & $F_1^{D\pi}$ & $F_1^{DK}$ & $A_0^{D\rho}$ & $A_0^{DK^*}$ \\
 \hline
 $F(0)$ & 0.666 & 0.739 & 0.666 & 0.739 & 0.74 & 0.78 \\
 $\alpha$ & 0.21 & 0.30 & 0.24 & 0.33 & 0.36 & 0.24\\
\end{tabular}
\end{ruledtabular}
\end{table}

\begin{table}[t]
\caption{The extracted parameters $a_1$ and $a_2$. }
\begin{ruledtabular} \label{tab:a12}
\begin{tabular}{l c c c c c }
 & &  \multicolumn{2}{c}{$D\to\ov K^*\pi$}
 &   \multicolumn{2}{c}{$D\to \ov K\rho$} \\ \cline{3-4} \cline{5-6}
\raisebox{2.0ex}[0cm][0cm]
 & \raisebox{2.0ex}[0cm][0cm]{$D\to \ov K\pi$} & A & S &A1 & S1 \\
 \hline
 $|a_1|$ & $1.22\pm0.02$ &  $1.32\pm0.05$ & $1.43\pm0.06$ & $1.38\pm0.06$ & $1.29\pm0.17$\\
 $|a_2|$ & $0.82\pm0.02$ &  $0.87\pm0.05$ &  $0.95\pm0.05$& $1.06\pm0.12$ & $0.82\pm0.27$\\
 $|a_2/ a_1|$ & $0.67\pm0.02$ &  $0.66\pm0.05$ & $0.67\pm0.05$ & $0.77\pm0.09$ & $0.63\pm0.22$\\
 Arg($a_2/a_1$) & $-(152 \pm 1)^\circ$ & $-(162 \pm 3)^\circ$ &
$-(164 \pm 3)^\circ$ & $(164 \pm 23)^\circ$ & $(177 \pm 15)^\circ$
\end{tabular}
\end{ruledtabular}
\end{table}

In order to extract the parameters $a_{1,2}$ we need to know the form factors and their $q^2$ dependence. There exist many model and lattice calculations for $D$ to $\pi,K$ transition form factors. In this work we shall use the following parametrization for form-factor $q^2$ dependence \cite{Melikhov}
\be \label{eq:FFDP}
F(q^2)={F(0)\over (1-q^2/m^2_*)(1-\alpha q^2/m_*^2)}
\en
with $m_*$ being a pole mass.  Specifically, $m_*=m_{D_s^*},~m_{D_s},~ m_{D^*},~m_{D}$ for form factors $F_{1,0}^{DK}$, $A_0^{DK^*}$, $F_{1,0}^{D\pi}$ and $A_0^{D\rho}$, respectively. The inputs for the parameters $F(0)$ and $\alpha$ are summarized in Table \ref{tab:FF}. Form factors for $D$ to $\pi$ and $K$ transitions are taken from the recent CLEO-c measurements of $D$ meson semileptonic decays to $\pi$ and $K$ mesons \cite{CLEO:FF}, while $D\to \rho,K^*$ transition form factors from \cite{Melikhov} with some modification. For decay constants we use $f_\pi=130.4$ MeV, $f_K=155.5$ MeV \cite{PDG},
$f_\rho=216$ MeV and $f_{K^*}=220$ MeV \cite{BallfV}.

The extracted parameters $a_{1,2}$ are shown in Table~\ref{tab:a12}, where we have used solutions A and S for the topological amplitudes $T_P, C_V$ and $\tilde T_P$, $\tilde C_V$, respectively, and the preferred solution A1 for $T_V, C_P$ and S1 for $\tilde T_V$, $\tilde C_P$. Due to final-state interactions, the magnitude of $a_2$ is substantially larger than the benchmark value 0.529\,, while the magnitude of $a_1$ is of order $1.3\sim1.4$ for $D\to \bar K^*\pi$ and $D\to \bar K\rho$ decays.

As noticed in passing, solutions of $C_P$ with magnitude smaller than $T_V$ (solutions A' and S' in Table \ref{tab:TV}) are ruled out since the predicted rates for SCS decays $D^0\to \pi^0\phi$ and $D^+\to\pi^+\phi$ are too small compared to experiments.  This can also be seen from the extracted ratio $|a_2/a_1|$ which is expected to be similar for all two-body hadronic $D$ decays. For solutions A' and S' with smaller $C_P$, the ratio $|a_2/a_1|$ is 0.38 and deviates substantially from the value of 0.68 obtained from $D\to\bar K\pi$ decays.  Notice that although $|C_P|>|T_V|$, the extracted parameters look normal, namely, $|a_1|>|a_2|$.  Among solutions A1$-$A6 and S1$-$S5, it is natural to select solutions A1 and S1 since the magnitude of their $a_2$ and the ratio $|a_2/a_1|$ are closer to those in solutions A and S (see Table \ref{tab:a12}).  As stressed before, a large $T_P$ is definitely needed in order to accommodate the data of $D_s^+\to\rho^+\eta$.

\subsection{SU(3) breaking and final-state interactions}

As discussed before, based on the exact flavor SU(3) symmetry and our best solutions to the topological amplitudes extracted from CF charm decays, the predicted branching fractions of SCS $D$ decays are too large for $\pi^+\pi^-$, $\pi^0\pi^0$ and too small for $K^+K^-$, $\pi^+\eta$ and $\pi^+\eta'$.  In the following, we shall first examine SU(3) breaking effects in color-allowed and color-suppressed tree amplitudes within the factorization approach (for an earlier study of SU(3) violation in charm decays, see \cite{CC94}). It turns out that while part of the SU(3) breaking effects can be accounted for by SU(3) symmetry violation manifested in $T$ and $C$ amplitudes,
in some cases such as the ratio $R=\Gamma(D^0\to K^+K^-)/\Gamma(D^0\to\pi^+\pi^-)$, SU(3) breaking alone in spectator amplitudes does not suffice to account for  $R$.
This calls for the consideration of SU(3) violation in the $W$-exchange amplitudes.

\subsubsection{SU(3) breaking}

To illustration the effect of SU(3) symmetry violation in external and internal $W$-emission amplitudes, we consider the following modes: $D^+\to\pi^+\pi^0,\pi^+\eta$, $\pi^+\eta'$  and  $D^0\to K^+K^-,~\pi^+\pi^-$

\vskip 0.3cm\noindent\underline{$D^+\to\pi^+\pi^0$}

Since the isospin of the $\pi^+\pi^0$ state is $I=2$, this channel does not have nontrivial final-state interactions. Indeed, it does not receive weak annihilation contributions. This means that the short-distance approach should suffice to describe this decay.  The factorizable amplitudes read
\be \label{eq:Amppipi}
T'_{\pi\pi} &=& {G_F\over\sqrt{2}}\,a_1 f_\pi(m_D^2-m_\pi^2)F_0^{D\pi}(m_\pi^2), \non \\
C'_{\pi\pi} &=& {G_F\over\sqrt{2}}\,a_2 f_\pi(m_D^2-m_\pi^2)F_0^{D\pi}(m_\pi^2).
\en
Hence,
\be
{T'_{\pi\pi}\over T} = {m_D^2-m_\pi^2\over m_D^2-m_K^2}\,{F_0^{D\pi}(m_\pi^2)\over F_0^{DK}(m_\pi^2)}, \qquad
{C'_{\pi\pi}\over C} = {f_\pi\over f_K}\,{F_0^{D\pi}(m_\pi^2)\over F_0^{D\pi}(m_K^2)},
\en
where use of Eq. (\ref{eq:Ampfac1}) has been made.
Numerically, we find $T'_{\pi\pi}/T=0.96$ and $C'_{\pi\pi}/C=0.78$.
Therefore,
\be
\A(D^+\to\pi^+\pi^0)={1\over\sqrt{2}}V_{cd}^*V_{ud}(0.96\,T+0.78\,C).
\en
Because of the less destructive interference between color-allowed and color-suppressed amplitudes due to SU(3) breaking effects, $\B(D^+\to\pi^+\pi^0)$ is enhanced from $(0.89 \pm 0.10) \times 10^{-3}$ to $(0.96 \pm 0.04) \times 10^{-3}$, in better agreement with the experiment.

\vskip 0.3cm\noindent\underline{$D^+\to\pi^+\eta^{(')}$}

The relevant factorizable amplitudes here are
\be
T'_{\pi\eta_q} &=& {G_F\over\sqrt{2}}\,a_1 f_\pi(m_D^2-m_{\eta_q}^2)F_0^{D\eta_q}(m_\pi^2), \non \\
C'_{\pi\eta_q} &=& {G_F\over\sqrt{2}}\,a_2 f_{q}(m_D^2-m_\pi^2)F_0^{D\pi}(m_{\eta_q}^2), \non \\
C'_{\pi\eta_s} &=& {G_F\over\sqrt{2}}\,a_2 f_{s}(m_D^2-m_\pi^2)F_0^{D\pi}(m_{\eta_s}^2),
\en
where $f_q$, $f_s$ are the decay constants of $\eta_q$ and $\eta_s$, respectively.
The masses of $\eta_q$ and $\eta_s$ read \cite{FKS}
\be
m_{\eta_q}^2 &=& {\sqrt{2}\over f_q}\la 0|m_u\bar ui\gamma_5u+m_d\bar di\gamma_5d|\eta_q\ra+{\sqrt{2}\over f_q}\la 0|{\alpha_s\over 4\pi}G\tilde G|\eta_q\ra\approx m_\pi^2+ {\sqrt{2}\over f_q}\la 0|{\alpha_s\over 4\pi}G\tilde G|\eta_q\ra \non \\
m_{\eta_s}^2 &=& {2\over f_s}\la 0|m_s\bar si\gamma_5s|\eta_s\ra+{1\over f_s}\la 0|{\alpha_s\over 4\pi}G\tilde G|\eta_s\ra\approx 2m_K^2-m_\pi^2+ {1\over f_s}\la 0|{\alpha_s\over 4\pi}G\tilde G|\eta_s\ra,
\en
where contributions to their masses from the gluonic anomaly have been included. We shall use the parameters extracted from a phenomenological fit \cite{KLOE,FKS}: $\phi=(40.4\pm0.6)^\circ$ and
\be
&& {1\over \sqrt{2}f_q}\la 0|{\alpha_s\over 4\pi}G\tilde G|\eta_q\ra=0.265\pm0.010, \non \\
&& { \la 0|{\alpha_s\over 4\pi}G\tilde G|\eta_q\ra\over \sqrt{2}\la 0|{\alpha_s\over 4\pi}G\tilde G|\eta_s\ra}={f_s\over f_q}=1.352\pm0.007, \quad f_q=f_\pi.
\en
Assuming that the unknown form factor $F_0^{D\eta_q}$ is the same as $F_0^{D\pi}$, the decay amplitudes are modified to
\be
\A(D^+\to\pi^+\eta) &=&{1\over\sqrt{2}}V_{cd}^*V_{ud}(0.85\,T+0.93\,C+2A)\cos\phi-V_{cs}^*V_{us}1.28\, C\sin\phi, \non \\
\A(D^+\to\pi^+\eta') &=&{1\over\sqrt{2}}V_{cd}^*V_{ud}(0.85\,T+0.93\,C+2A)\sin\phi+V_{cs}^*V_{us}1.28\, C\cos\phi,
\en
where we have assumed $A'=A$. It follows that $\B(D^+\to\pi^+\eta)=(2.68 \pm 0.29)\times 10^{-3}$ [$(3.54 \pm 0.21)\times 10^{-3}$] and $\B(D^+\to\pi^+\eta')=(5.03 \pm 0.32)\times 10^{-3}$ [$(4.68\pm0.30)\times 10^{-3}$], where the data are shown in the squared brackets. Hence the discrepancy between theory and experiment is substantially improved. Presumably, a better agreement will be achieved if SU(3) violation in the $W$-annihilation amplitude is included.  For example, if $A'=0.7A$ is taken, we will have $\B(D^+\to\pi^+\eta)=(3.02 \pm 0.19)\times 10^{-3}$ and $\B(D^+\to\pi^+\eta')=(4.69 \pm 0.21)\times 10^{-3}$, bringing our predictions closer to the observed data.

\vskip 0.3cm\noindent\underline{$D^0\to\pi^+\pi^-,K^+K^-$}

Experimentally, the rate of $D^0\to K^+K^-$ is larger than that of $D^0\to \pi^+\pi^-$ by a factor of 2.8, while they should be the same in the SU(3) limit. This is a long-standing puzzle since SU(3) symmetry is expected to be broken at the level of 30\%. Without the inclusion of SU(3) breaking effects in the topological amplitudes, we see from Table~\ref{tab:CSPP} that the predicted rate of $K^+K^-$ is even smaller than that of $\pi^+\pi^-$ due to less phase space available to the former.  In the factorization approach we have
\be
{T'_{KK}\over T'_{\pi\pi}}={f_K\over f_\pi}\,{m_D^2-m_K^2\over m_D^2-m_\pi^2}\,{F_0^{DK}(m_K^2)\over F_0^{D\pi}(m_\pi^2)}.
\en
Using the form factor parametrization given by Eqs. (\ref{eq:FFDP}) and input parameters given in Table \ref{tab:FF}, and assuming no SU(3) symmetry breaking in $E$ so that $E'_{KK}=E'_{\pi\pi}=E$, we obtain $T'_{KK}/T'_{\pi\pi}=1.32$, implying that $\B(D^0\to K^+K^-) = (3.4 \pm 0.1) \times 10^{-3}$ and $\B(D^0\to\pi^+\pi^-)= (2.1 \pm 0.1) \times 10^{-3}$. Therefore, SU(3) symmetry breaking alone in $T'$ is not adequate to resolve the puzzle. This calls for the consideration of SU(3) flavor symmetry violation in the $W$-exchange amplitudes.  We will come back to this issue in the next subsection.

\subsubsection{Final-state interactions}

We learn from the above few examples that when SU(3) breaking effects in color-allowed and color-suppressed amplitudes are included, the discrepancy between theory and experiment for SCS decays will be significantly reduced. However, in order to get a better agreement, violation of SU(3) symmetry in weak annihilation ($W$-exchange and $W$-annihilation) must be taken into account. This is particularly true for the ratio $R=\Gamma(D^0\to K^+K^-)/\Gamma(D^0\to \pi^+\pi^-)$; SU(3) violation in spectator amplitudes accounts for only half of the ratio $R$.

\begin{figure}[t]
\begin{center}
\includegraphics[width=0.80\textwidth]{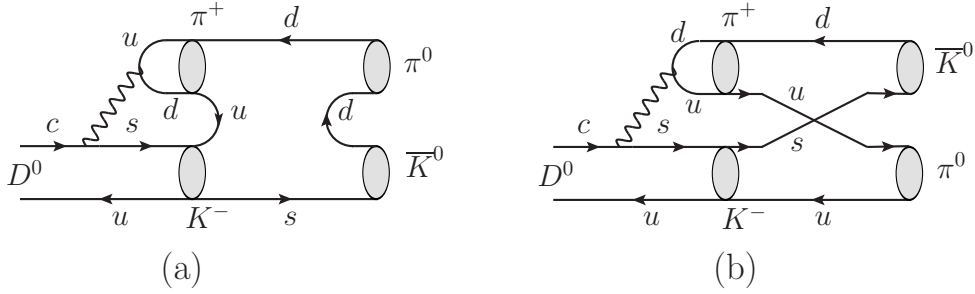}
\vspace{0.0cm}
\caption{Contributions to $D^0\to \ov K^0\pi^0$ from
    the color-allowed weak decay $D^0\to K^-\pi^+$ followed by a
    resonant-like rescattering (a) and quark exchange (b). While (a)
    has the same topology as the $W$-exchange graph, (b) mimics
    the color-suppressed internal $W$-emission graph.} \label{fig:DFSI} \end{center}
\end{figure}

Under the factorization hypothesis, the factorizable $W$-exchange and $W$-annihilation amplitudes are proportional to $a_2$ and $a_1$, respectively. \footnote{In the QCD factorization (QCDF) approach \cite{BBNS}, $E$ and $A$ are proportional to the Wilson coefficients $c_1$ and $c_2$, respectively, and hence $A=(c_2/c_1)E$ \cite{Lai,Gao}. However, the charm quark is not heavy enough to justify the use of QCDF in charm decays.} They are suppressed due to the smallness of the form factor at large $q^2=m_D^2$.  This corresponds to the so-called helicity suppression. At first glance, it appears that the factorizable weak annihilation amplitudes are too small to be consistent with experiment at all.  However, in the diagrammatic approach here, the topological amplitudes $C,~E,~A$ can receive contributions from the tree amplitude $T$ via final-state rescattering, as illustrated in Fig.~\ref{fig:DFSI} for $D^0\to \bar K^0\pi^0$: Fig.~\ref{fig:DFSI}(a) has the same topology as the $W$-exchange diagram $E$, while \ref{fig:DFSI}(b) mimics the internal $W$-emission amplitude $C$. Therefore, even if the short-distance $W$-exchange vanishes, a long-distance $W$-exchange can be induced via inelastic FSIs. Historically, it was first pointed out in \cite{Donoghue} that rescattering effects required by unitarity can produce the reaction $D^0\to\ov K^0\phi$, for example, even in the absence of the $W$-exchange diagram. Then it was realized that this rescattering diagram belongs to the generic $W$-exchange topology \cite{CC87}.

There are several different forms of FSIs: elastic scattering and inelastic scattering such as quark exchange, resonance formation, etc.  For charm decays, it is expected that the long-distance weak annihilation is dominated by resonant FSIs.  That is, the resonance formation of FSI via $q\bar q$ resonances is usually the most important one due to the fact that an abundant spectrum of resonances is known to exist at energies close to the mass of the charmed meson. Indeed, the sizable magnitude of $E$ and its large phase determined from experiment are suggestive of nearby resonance effects. (As noted in \cite{Chenga1a2}, weak annihilation in $VP$ systems receives little contributions from resonant FSIs. We shall show below that the $W$-annihilation amplitude in $D_s^+\to\pi^+\omega$ arises from final-state rescattering via quark exchange.)  A direct calculation of the resonant FSI diagram is subject to many theoretical uncertainties.  Nevertheless, as emphasized in \cite{Zen,Weinberg}, most of the properties of resonances follow from unitarity alone, without regard to the dynamical mechanism that produces the resonance. Consequently, as shown in \cite{Zen,Chenga1a2}, the effect of resonance-induced FSIs can be described in a model-independent manner in terms of the mass and width of the nearby resonances. It was found that weak annihilation amplitudes are modified by resonant FSIs as (see {\it e.g.}, \cite{Chenga1a2})
 \be \label{E}
 E = e+(e^{2i\delta_r}-1)\left(e+{T\over 3}\right), \qquad
  A = a+(e^{2i\delta_r}-1)\left(a+{C\over 3}\right),
 \en
with
  \be \label{phase}
e^{2i\delta_r}=1-i\,{\Gamma\over m_D-m_R+i\Gamma/2},
 \en
where the $W$-exchange amplitude $E$ and $W$-annihilation $A$ before resonant FSIs are denoted by $e$ and $a$, respectively. Therefore, resonance-induced FSIs amount to modifying the $W$-exchange amplitude and leaving the other quark-diagram amplitudes $T$ and $C$ intact. We thus see that even if the short-distance weak annihilation vanishes (i.e. $e=0$ and $a=0$), as commonly asserted, a long-distance $W$-exchange ($W$-annihilation) contribution still can be induced from the tree amplitude $T$ ($C$) via FSI rescattering in resonance formation.

\vskip 0.3cm\noindent\underline{$D^0\to\pi^+\pi^-,K^+K^-$ revisited}

\begin{figure}[t]
\begin{center}
\includegraphics[width=0.37\textwidth]{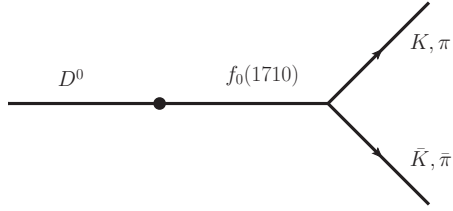}
\vspace{0.0cm}
\caption{Long-distance resonant contribution to the color-suppressed tree amplitude of $D^0\to K\bar K,\pi\bar\pi$ through the intermediate state $f_0(1710)$, where the blob stands for a transition due to weak interactions. This has the same topology as the $W$-exchange topological diagram.} \label{fig:f01710}
\end{center}
\end{figure}

We have discussed before that resonant FSIs usually give the most important contributions to the weak annihilation topology in the $PP$ modes. For $\Delta S=1$
$D\to \bar K\pi$ decays, there is a $J^P=0^+$ resonance
$K^*_0(1950)$ in the $s\bar d$ quark content with a mass $1945\pm
10\pm 20$ MeV and a width $201\pm 34\pm 79$ MeV \cite{PDG}. \footnote{The importance of the $K^*_0(1950)$ contribution to $D^0\to K^-\pi^+$ has been emphasized in \cite{GronauD}.}
Assuming
$e=0$ in Eq. (\ref{E}), we obtain
 \be
 E=1.68\times 10^{-6}\,{\rm exp}(i143^\circ)\,{\rm GeV},
 \en
 which is close to the ``experimental" value $E=(1.53^{+0.05}_{-0.06} )\times 10^{-6}\,{\rm exp}[i(122\pm1)^\circ]$ GeV.  Presumably, a non-vanishing short-distance $e$ will render the phase of $E$ in agreement with experiment.  For $\Delta S=0$ $D\to PP$ decays, there is a nearby scalar resonance $f_0(1710)$ which decays to $K\bar K$ and $\pi\bar\pi$, as depicted in Fig.~\ref{fig:f01710}. This long-distance contribution has the same topology as the $W$-exchange diagram.

Recent lattice and phenomenological studies indicate that $f_0(1710)$ is dominated by the scalar glueball component \cite{Chen,CCL}.  If $f_0(1710)$ is primarily a scalar glueball $G$, it is na{\"i}vely expected that $\Gamma(G\to\pi\bar\pi) / \Gamma(G\to K\bar K) \approx 0.9$ after phase space correction due to the flavor independent coupling of $G$ to $PP$. However, experimentally there is a relatively large suppression of $\pi\bar\pi$ production relative to $K\bar K$ in $f_0(1710)$ decays. The ratio $\Gamma(f_0(1710)\to \pi\bar\pi) / \Gamma(f_0(1710)\to K\bar K)$ is measured to be $0.41^{+0.11}_{-0.17}$ by BES from $J/\psi \to \gamma(K^+K^-,\pi^+\pi^-)$ decays \cite{BESgammapipi}.  To explain the large disparity between $\pi\bar\pi$ and $K\bar K$ production in scalar glueball decays, it was first noticed by Carlson et al. \cite{Carlson}, by Cornwall and Soni \cite{Cornwall} and  vitalized recently by Chanowitz \cite{Chanowitz} that a pure scalar glueball cannot decay into a quark-antiquark pair in the chiral limit, {\it i.e.}, $\A(G\to q\bar q)\propto m_q$.  Since the current strange quark mass is an order of magnitude larger than $m_u$ and $m_d$, decay to $K\bar K$ is largely favored over $\pi\bar\pi$.  However, chiral suppression for the ratio $\Gamma(G\to \pi\bar\pi)/\Gamma(G\to K\bar K)$ at the hadron level should not be so strong as the current quark mass ratio $m_u/m_s$.  It has been suggested \cite{Chao} that $m_q$ should be interpreted as the scale of chiral symmetry breaking since chiral symmetry is broken not only by finite quark masses but is also broken spontaneously.

In short, if the glueball component of $f_0(1710)$ is dominant, then this will explain qualitatively why the $D^0$ meson decays to $K^+K^-$ more copiously then $\pi^+\pi^-$ through the resonant FSIs.

\vskip 0.3cm\noindent\underline{$D^0\to K^0\bar K^0$}

\begin{figure}[t]
\begin{center}
\includegraphics[width=0.7\textwidth]{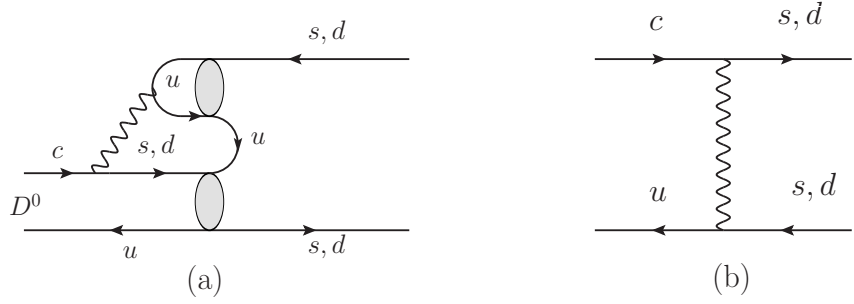}
\vspace{0.0cm}
\caption{Possible $D^0-f_0(1710)$ transition from (a)  intermediate states such as $K^+K^-,\pi^+\pi^-$ etc., and (b) $W$-exchange. } \label{fig:D0f01710} \end{center}
\end{figure}

It is known that the decay $D^0\to K^0\bar K^0$ proceeds only through the $W$-exchange mechanism and vanishes under SU(3). Since the short-distance $W$-exchange is expected to be small and furthermore subject to SU(3) cancellation, this mode can only occur through long-distance $W$-exchange induced by final-state rescattering \cite{Pham}.  The nearby pole contribution from $f_0(1710)$ will contribute to the decay $D^0\to K^0\bar K^0$.  However, in contrast to $D^0\to K^+K^-$, this channel is prohibited under exact SU(3) symmetry.  To see this, Fig.~\ref{fig:D0f01710} shows some possible weak transition between $D^0$ and $f_0(1710)$.  It is clear that for $K^0\bar K^0$ production, both $s$ and $d$ quarks get involved and compensate each other in SU(3) limit due to the cancellation of CKM matrix elements, while only the $s$ quark gets involved in $K^+K^-$ production. It has been shown in \cite{CCL} that in the limit of SU(3) symmetry, $f_0(1710)$ is composed of a scalar gluonium and a small amount of SU(3) singlet.  Consequently, the $D^0\!-\!f_0(1710)$ weak transition for $K^0\bar K^0$ production indeed vanishes when SU(3) flavor symmetry is exact.

\subsection{Annihilation amplitudes $A_P$ and $A_V$ from final-state rescattering}

As discussed before, we have some difficulties in extracting and understanding the $W$-annihilation amplitudes $A_P$ and $A_V$ in $D\to VP$ decays.  The topological amplitude expressions of $D_s^+\to \pi^+\rho^0$ and $D_s^+\to\pi^+\omega$ are given by
\be
\A(D_s^+\to \pi^+\rho^0) &=& {1\over\sqrt{2}}V_{cs}^*V_{ud}(A_V-A_P), \non \\
\A(D_s^+\to \pi^+\omega) &=& {1\over\sqrt{2}}V_{cs}^*V_{ud}(A_V+A_P).
\en
Na{\"i}vely it is expected that $A_P=-A_V$.  The argument goes as follows. The direct $W$-annihilation contributions via $c\bar s\to W\to u\bar d$ are not allowed in $D_s^+\to\pi^+\omega,\,\rho^+\eta,\,\rho^+\eta'$ decays since the $(u\bar d)$ has zero total angular momentum and hence it has the quantum number of $\pi^+$.  Therefore, $G(u\bar d)=-$ and the final states should have an odd $G$-parity.  Since the $G$-parity is even for $\omega\pi^+$ and odd for $\pi^+\rho^0$, it follows that the former does not receive any direct $W$-exchange contribution.  Can one induce $D_s^+\to\pi^+\omega$ from resonant FSIs?  The answer is no because the $J=0,~I=1$ meson resonance made from a quark-antiquark pair $u\bar d$ has an odd $G$ parity.  As stressed in \cite{Lipkin}, the even-$G$ state $\pi^+\omega$ (also $\rho\eta$ and $\rho\eta'$) does not couple to any single meson resonances, nor to the state produced by the $W$-annihilation diagram with no gluons emitted by the initial state before annihilation.  Indeed, a general consideration of resonant FSIs gives the relations \cite{Zen,Chenga1a2}
\be
 A_P^r+A_V^r &=& a_P+a_V, \non \\
 A_P^r-A_V^r &=& a_P-a_V+(e^{2i\delta_r}-1)\left(a_P-a_V+{1\over
 3}(C_P-C_V)\right),
\en
with the superscript $r$ denoting the annihilation amplitude arising from resonant FSIs.  The above relation shows that $A_P+A_V$ does not receive any $\bar qq'$ resonance ({\it e.g.}, the $0^-$ resonance $\pi(1800)$) contributions.  Since the above $G$-parity argument implies $a_V= -a_P$, the decay $D_s^+\to\pi^+\omega$ is forbidden, whereas $D_s^+\to\pi^+\rho^0$ receives both factorizable and resonance-induced $W$-annihilation contributions.  Experimentally, however, it is the other way around: $\B(D_s^+\to\pi^+\omega) = (2.1\pm 0.9)\times 10^{-3}$, while $\B(D_s^+\to\pi^+\rho^0)$ is not seen.

\begin{figure}[t]
\begin{center}
\includegraphics[width=0.80\textwidth]{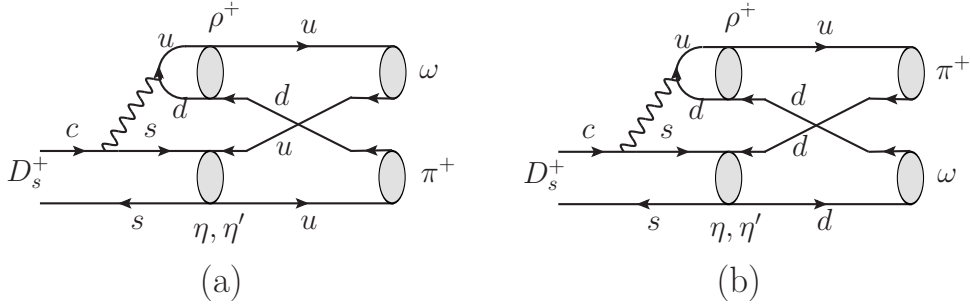}
\vspace{0.0cm}
\caption{Long-distance contributions to $D_s^+\to \pi^+\omega$ from the weak decay $D_s^+\to\rho^+\eta^{(')}$ followed by quark exchange.  Plots (a) and (b) have the same topology as $A_V$ and $A_P$, respectively. Owing to $G$-parity conservation, $\rho^+\eta^{(')}$ cannot rescatter into $\pi^+\rho^0$. } \label{fig:DstoPiOmega} \end{center}
\end{figure}

To resolve the above-mentioned puzzle, we note that there are long-distance final-state rescattering contributions to $D_s^+\to \pi^+ \omega$ allowed by $G$-parity conservation.  A nice example is the contribution from the weak decay $D_s^+\to\rho^+\eta^{(')}$ followed by quark exchange (Fig.~\ref{fig:DstoPiOmega}). The rescattering of $\rho^+\eta^{(')}$ into $\pi^+\rho^0$ is prohibited by the $G$-parity selection rule. Consequently, $A_P+A_V=A_P^e+A_V^e=2A_P^r$, where the superscript $e$ indicates final-state rescattering via quark exchange, and $A_P-A_V=A_P^r-A_V^r$. Since $D_s^+\to\rho^+\eta$ has the largest rate among the CF $D_s^+\to VP$ decays, \footnote{As discussed before, from the theoretical point of view, $\B(D_s^+\to\rho^+\eta')$ is at most of order 3\%. } it is conceivable that $D_s^+\to \pi^+\omega$ can be produced via FSIs at the $10^{-3}$ level as its branching fraction. Other processes such as the weak decays $D_s^+\to \bar K^{(*)0}K^{(*)+}$ followed by $\bar K^{(*)0}K^{(*)+}\to \pi^+\omega$ have been discussed in \cite{Kamal,Fajfer}. However, this rescattering process does not involve quark exchange. This decay may also proceed through pre-radiation of the $\omega$. For instance, the $D_s^+$ meson can dissociate into two meson states such as $D^{(*)0}K^{(*)+}$ and $D^{(*)+}K^{(*)0}$ which rescatter strongly to $(c\bar s)\omega$ while the virtual $c\bar s$ state decays weakly to $\pi^+$ \cite{Gronau2}.

As pointed out in \cite{Gronau}, the $\omega$ meson can be produced from the decay $D_s^+\to\pi^+\phi$ followed by $\omega-\phi$ mixing. This will be possible if $\phi$ is not a pure $s\bar s$ state and contains a tiny $q\bar q$ component. Neglecting isospin violation and the admixture with the $\rho^0$ meson, one can parametrize the $\omega$--$\phi$ mixing in terms of an angle $\delta$ such that the physical $\omega$ and $\phi$ are related to the ideally mixed states $\omega^I \equiv (u \bar u + d \bar d)/\sqrt{2}$ and $\phi^I \equiv s \bar s$ by
\begin{eqnarray}\label{mixing}
\left(
\begin{array}{c} \omega \\ \phi \end{array} \right) = \left(
\begin{array}{c c} \cos \delta & \sin \delta \\ - \sin \delta & \cos
\delta
\end{array} \right)
\left( \begin{array}{c} \omega^I \\ \phi^I \end{array} \right),
\end{eqnarray}
and the mixing angle is about $|\delta| \sim 3.3^\circ$ \cite{Benayoun:1999fv} (see \cite{KLOE} for the latest determination of $\delta$). Therefore, the production of $\pi^+\omega$ through $\omega-\phi$ mixing is expected to be
\begin{eqnarray}
\B(D_s^+\to \pi^+\omega)_{\rm \omega-\phi~ mixing}=\B
(D_s^+\to \pi^+\phi)\left({p_c(\pi\omega)\over p_c(\pi\phi)}\right)^3\tan^2\delta \sim 2.2\times 10^{-4},
\end{eqnarray}
which is too small by one order of magnitude. Hence, it is not the dominant mechanism for the $D_s^+\to \pi^+\omega$ decay.

\begin{figure}[t]
\begin{center}
\includegraphics[width=0.80\textwidth]{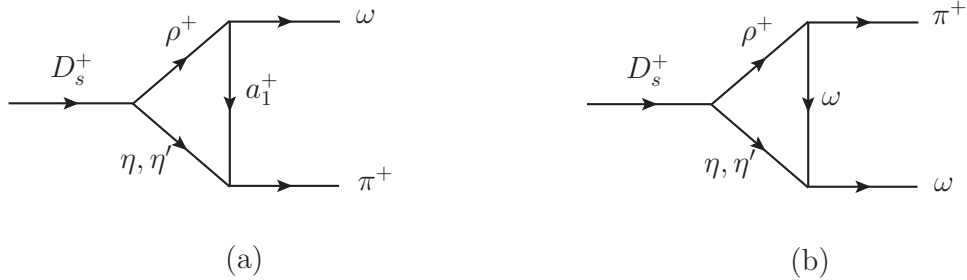}
\vspace{0.0cm}
\caption{ Manifestation of Fig. \ref{fig:DstoPiOmega} as
the long-distance {\it t}-channel  contributions to the $W$-annihilation amplitude of $D_s^+\to\pi^+\omega$.} \label{fig:Dspiomega} \end{center}
\end{figure}

Since FSIs are nonperturbative in nature, in principle it is extremely difficult to calculate their effects. It is customary to evaluate the final-state rescattering contribution, Fig.~\ref{fig:DstoPiOmega}, at the hadron level manifested in Fig.~\ref{fig:Dspiomega}. One of the diagrams, namely, Fig.~\ref{fig:Dspiomega}(b), has been evaluated in \cite{Fajfer}. As stressed in \cite{CCSfsi}, the calculation of the meson-level Feynman diagrams such as Fig.~\ref{fig:Dspiomega} involves many theoretical uncertainties.  If one na{\"i}vely calculates the diagram, one will obtain an answer which does not make sense in the context of perturbation theory since the contributions become so large that perturbation is no longer trustworthy.  Moreover, because the $t$-channel exchanged particle is not on-shell, a form-factor cutoff must be introduced to the vertex to render the whole calculation meaningful. We will leave the calculation of Fig.~\ref{fig:Dspiomega} to a separate publication.

Finally, we would like to point out that the study of the decays $D_s^+\to (\rho^+,a_1^+,a_0^+)\omega$ will help understand the annihilation mechanisms in $D_s^+$ decays. Unlike the $\pi^+\omega$ mode, $D_s^+\to \rho^+\omega$ can be induced by resonant FSIs and hence it is anticipated to have a larger rate. The recent CLEO measurements of $D_s^+$ exclusive decays involving an $\omega$ show that the branching fractions of $D_s^+\to (\pi^+\pi^0,~\pi^+\pi^+\pi^-,~\pi^+\eta) \omega$ are $(2.78\pm0.70)\%$, $(1.58\pm0.46)\%$ and $(0.85\pm0.54)\%$, respectively \cite{CLEO:omega}. Assuming the validity of the narrow width approximation, $\B(a_0^+(980)\to \pi^+\eta)=1$ and $\B(a_1^+(1260)\to (3\pi)^+)=100\%$ in conjunction with the assumption of $\B(a_1^+\to\pi^+\pi^+\pi^-)=\B(a_1^+\to\pi^+\pi^0\pi^0)$, the branching fractions of $D_s^+\to a_0^+(980)\omega$ and $D_s^+\to a_1^+(1260)\omega$ are ready to obtain.

\subsection{Comments on $D\to VV$ and $D\to (S,A,T)(P,V)$ decays}

\vskip 0.3cm\noindent\underline{$D\to VV$}

A handful number of $VV$ modes, such as $D^+\to \bar K^{*0}\rho^+$, $D^0\to \bar K^{*0}\rho^0,~K^{*-}\rho^+,~K^{*0}\bar K^{*0},~\rho^0\rho^0,~\phi\rho^0$, $\bar K^{*0}\omega$ and $D_s^+\to \rho^+\phi,~\rho^+\omega,~K^{*+}\bar K^{*0}$, have been observed.  These decays have a richer structure as they have three partial-wave or helicity states.  Na{\"i}vely it is expected from the factorization hypothesis that the longitudinal polarization is dominant. However, the Mark III measurement \cite{MarkIII} has indicated that the branching fraction of the decay $D^0\to \bar K^{*0}\rho^0$ is already saturated by the transverse polarization state, while longitudinal and transverse polarizations are comparable in $D^0\to K^{*-}\rho^+$ (see, {\it e.g.}, \cite{Kamal99} for a detailed discussion).  A Dalitz-plot study of four-body $D$ decays will be helpful to resolve the puzzle.

\vskip 0.3cm\noindent\underline{$D\to (S,A,T)(P,V)$}

Thanks to the powerful Dalitz plot analyses of multi-body decays of $D$ mesons, the structure of many resonances can be probed.  Besides the vector meson resonances, many parity-even states such as the scalar mesons $a_0(980), f_0(980),f_0(1370), f_0(1500), f_0(1710),a_0(1450),K_0^*(1430)$, the axial-vector mesons $a_1(1260),K_1(1270),K_1(1400)$ and the tensor mesons $f_2(1270),a_2(1320),K_2^*(1430)$ have been studied.  Branching fractions of quasi-two-body decays $D\to SP,~ VP,~TP$ with $S,A,T$ standing for scalar, axial-vector and tensor mesons, respectively, can be extracted from the Dalitz analyses of three-body decays of $D$ mesons.  The study of $D\to SP$ opens a new avenue to the understanding of the light scalar meson spectroscopy, recalling that the underlying structure of light scalar mesons $\sigma, \kappa, a_0(980)$ and $f_0(980)$ is not well established experimentally and theoretically.

The topological amplitudes for $D\to SP,AP,TP$ decays have been discussed in \cite{ChengSP,ChengAP,ChengTP}.  There are several new features.  First, one generally has two sets of distinct external $W$-emission and internal $W$-emission diagrams, depending on whether the emitted particle is a party-even meson or a parity-odd one.  Let us denote the primed amplitudes $T'$ and $C'$ for the case when the emitted meson is a parity-even one.  Secondly, because of the smallness of the decay constants of parity-even mesons except for the $^3P_1$ axial-vector state, it is expected that $|T'|\ll |T|$ and $|C'|\ll |C|$. This feature can be tested experimentally.  Thirdly, since $K^*_0$ and the light scalars $\sigma,~\kappa,~f_0,~a_0$ fall into two different SU(3) flavor nonets, one cannot apply SU(3) symmetry to relate the topological amplitudes in $D^+\to f_0\pi^+$ to, for example, those in $D^+\to \ov K^{*0}_0\pi^+$.  Finally, we notice that the new data of quasi-two-body decays have been accumulated in the past years to the extent that they warrant a serious theoretical study.

\section{Summary and Conclusion}

We have studied in this work the two-body hadronic charmed meson decays, including both the $PP$ and $VP$ modes.  The latest experimental data are first analyzed in the diagrammatic approach.  The magnitudes and strong phases of the flavor amplitudes are extracted from the Cabibbo-favored (CF) decay modes using $\chi^2$ minimization.  The best-fitted values are then used to predict the branching fractions of the singly-Cabibbo-suppressed (SCS) and doubly-Cabibbo-suppressed decay modes in the flavor SU(3) symmetry limit.  In doing so, we have observed SU(3) breaking effects more significant than the two-body hadronic bottom meson decays.

In Section~\ref{sec:topology}, we find from the CF $PP$ modes that
\be
&& T=3.14\pm0.06, \qquad\qquad\qquad\quad
C=(2.61\pm0.08)\,e^{-i(152\pm1)^\circ}, \non \\
&&  E=(1.53^{+0.07}_{-0.08})\,e^{i(122\pm2)^\circ},
\qquad\quad  A=(0.39^{+0.13}_{-0.09})\,e^{i(31^{+20}_{-33})^\circ}
\en
for the $\eta$-$\eta'$ mixing angle $\phi=40.4^\circ$ inferred by KLOE \cite{KLOE}.  Due to the larger branching fractions of ${\bar K}^0 \eta$ and $\pi^+\eta^{(\prime)}$ modes, the magnitudes of $T$ and $C$ quoted above are larger than previously obtained.  Also, the $T$ and $C$ amplitudes subtend a strong phase of about $150^\circ$.  Moreover, the $E$ and $A$ amplitudes are roughly perpendicular to each other.

In the case of $VP$ modes, we find two possible solutions for the $T_V$, $C_P$, and $E_P$ amplitudes, as given in Table~\ref{tab:TV}.  The SCS $VP$ modes $D^{0,+} \to \pi^{0,+} \phi$ favor the solution with a large $C_P$ (namely, solutions A and S in Table~\ref{tab:TV} depending on which formula is used to extract the invariant amplitude).  By assuming that $T_P$ and $T_V$ are relatively real, we then obtain several possible solutions for $T_P$, $C_V$, and $E_V$, as listed in Table~\ref{tab:TP}.  Again, only one of them (solutions A1 and S1, with the largest $T_P$) is in better agreement with the SCS $VP$ modes, such as the $D_s^+ \to \rho^+ \eta$ mode.  In this case, $|(T,C,E)_P| > |(T,C,E)_V|$.  The conclusion does not change much even if we make the relative phase between $T_P$ and $T_V$ a free parameter because it turns out to be very close to zero.  Finally, the $A_P$ and $A_V$ amplitudes cannot be completely determined based on currently available data.  Instead, we use the non-observation of $D_s^+ \to \pi^+ \rho^0$ to argue that they are of similar magnitudes, and use the measured rate of $D_s^+ \to \pi^+ \omega$ to show that they are much smaller in size than the color-suppressed ones.  However, this leads to a contradiction with data, as $\B(D_s^+ \to {\bar K}^0 K^{*+}) > \B(D_s^+ \to {\bar K}^{*0} K^+)$ while the former is dominated by $C_V$ and the latter by $C_P$.  So a full determination of the $A_P$ and $A_V$ amplitudes still await more precise data on the related modes. We conjecture that the currently quoted experimental results for both $D_s^+\to\bar K^0K^{*+}$ and $D_s^+\to \rho^+\eta'$ are overestimated and problematic.

In Section~\ref{sec:pheno}, we compare the sizes of color-allowed and color-suppressed tree amplitudes extracted in Section~\ref{sec:topology} with the effective parameters $a_1$ and $a_2$ defined in the factorization approach.  The extracted $a_2$, as given in Table~\ref{tab:a12}, is significantly larger than that obtained from short-distance calculations.  The ratio $|a_2/a_1|$ is more or less universal among the $D \to {\bar K} \pi$, ${\bar K}^* \pi$ and ${\bar K} \rho$ modes.  This feature allows us to discriminate between different solutions of topological amplitudes.  The relative strong phase extracted from the ${\bar K} \rho$ modes and that from the ${\bar K} \pi$ and ${\bar K}^* \pi$ modes are all in the vicinity of $180^\circ$, as expected from na{\"i}ve factorization.

Some Cabibbo-suppressed modes exhibit sizable violation of flavor SU(3) symmetry.  We find that part of the SU(3) breaking effects can be accounted for by SU(3) symmetry violation manifested in the color-allowed and color-suppressed tree amplitudes.  However, in other cases such as the ratio $R=\Gamma(D^0\to K^+K^-)/\Gamma(D^0\to\pi^+\pi^-)$, SU(3) breaking in spectator amplitudes is not sufficient to explain the observed value of $R$.  This calls for the consideration of SU(3) violation in the $W$-exchange amplitudes. Since weak annihilation topologies in $D\to PP$ decays are dominated by nearby scalar resonances via final-state rescattering (for example, the resonance $K_0^* (1950)$ helps explain the magnitude and phase of $E$), we argue that the long-distance resonant contribution through the intermediate state $f_0(1710)$ can naturally explain why $D^0$ decays more copiously to $K^+ K^-$ than $\pi^+ \pi^-$ through the $W$-exchange topology. This has to do with the dominance of the scalar glueball content of $f_0(1710)$ and the chiral-suppression effect in the decay of a scalar glueball into two pseudoscalar mesons. The same FSI through the $f_0(1710)$ pole contribution also explains the occurrence of $D^0\to K^0\bar K^0$.  However, in contrast to $D^0\to K^+K^-$, this channel vanishes under SU(3) symmetry owing to the cancelation of CKM matrix elements $V_{cs}^*V_{us}$ and $V_{cd}^*V_{ud}$ and the fact that $f_0(1710)$ is composed of a scalar gluonium and a small amount of SU(3) singlet in the SU(3) limit.

The CF decays $D_s^+\to \rho^+\pi^0,~\rho^0\pi^+,~\pi^+\omega,~\rho^+\omega$ proceed only through the $W$-annihilation topology.  Owing to the $G$-parity selection rule, $D_s^+\to\pi^+\omega$ does not receive contributions from the short-distance $W$-annihilation and resonant FSIs.  Nevertheless, it can proceed through the weak decays $D_s^+\to\rho^+\eta^{(')}$ followed by the final-state rescattering of $\rho^+\eta^{(')}$ into $\pi^+\omega$ through quark exchange.  In contrast, $D_s^+\to\rho^+\omega$ can be induced by resonant FSIs and hence it is anticipated to have a larger rate.

Finally we have made some brief remarks on other hadronic decays such as $D\to VV$ and $D\to (S,A,T)P$, the detailed investigation of which is left to a future work.

\section*{Acknowledgments}

We are grateful to B.~Bhattacharya for useful communications.  One of us (H.-Y.~C.) wishes to thank the hospitality of the Physics Department, Brookhaven National Laboratory.  This research was supported supported in part by the National Science Council of Taiwan, R.~O.~C.\ under Grant Nos.~NSC~97-2112-M-008-002-MY3, NSC~97-2112-M-001-004-MY3 and in part by the NCTS.

\newpage


\begin{thebibliography}{99}

\bibitem{Fuk}
M. Fukugita, T. Inami, N. Sakai, and S. Yazaki, \pl {\bf 72B}, 237
(1977); D. Tadi\'c and J. Trampeti\'c,  ibid. {\bf 114B}, 179
(1982); M. Bauer and B. Stech, ibid. {\bf 152B}, 380 (1985).

\bibitem{Buras}
A.J. Buras, J.-M. G\'erard, and R. R\"uckl, Nucl. Phys. B {\bf 268}, 16 (1986).

\bibitem{BS} B. Blok and M. Shifman, Sov. J. Nucl. Phys. {\bf
45}, 35, 301, 522 (1987).

\bibitem{CLEOPP09}
  H.~Mendez {\it et al.}  [CLEO Collaboration],
  arXiv:0906.3198 [hep-ex].


\bibitem{Wu}
  Y.~L.~Wu and M.~Zhong,
  Int.\ J.\ Mod.\ Phys.\  A {\bf 24}, 569 (2009).


\bibitem{Ryd}
  A.~Ryd and A.~A.~Petrov,
  arXiv:0910.1265 [hep-ph].

\bibitem{CC87} L.L. Chau and H.Y. Cheng, Phys. Rev. D {\bf 36}, 137 (1987); Phys. Lett. B
{\bf 222}, 285 (1989).

\bibitem{Chau} L.L. Chau, Phys. Rep. {\bf 95}, 1 (1983).

\bibitem{CC86} L.L. Chau and H.Y. Cheng, Phys. Rev. Lett. {\bf 56}, 1655 (1986).

\bibitem{Gronau94} M. Gronau, O.F. Hernandez, D. London, and J.L. Rosner,
  Phys. Rev. D {\bf 50} 4529 (1994); Phys. Rev. D {\bf 52} 6374 (1995)


\bibitem{Rosner99} J.L. Rosner, Phys. Rev. D {\bf 60}, 114026 (1999).

\bibitem{Chiang03}
  C.~W.~Chiang, Z.~Luo and J.~L.~Rosner,
  Phys.\ Rev.\  D {\bf 67}, 014001 (2003)
  [arXiv:hep-ph/0209272].

\bibitem{Wu04}
  M.~Zhong, Y.~L.~Wu and W.~Y.~Wang,
  Eur.\ Phys.\ J.\  C {\bf 32}, 191 (2004).

\bibitem{Wu05}
  Y.~L.~Wu, M.~Zhong and Y.~F.~Zhou,
  Eur.\ Phys.\ J.\  C {\bf 42}, 391 (2005)
  [arXiv:hep-ph/0405080].

\bibitem{RosnerPP08}
  B.~Bhattacharya and J.~L.~Rosner,
  Phys.\ Rev.\  D {\bf 77}, 114020 (2008)
  [arXiv:0803.2385 [hep-ph]].

\bibitem{RosnerVP}
  B.~Bhattacharya and J.~L.~Rosner,
  Phys.\ Rev.\  D {\bf 79}, 034016 (2009)
  [arXiv:0812.3167 [hep-ph]].

\bibitem{CLEOPP08}
  M.~Artuso {\it et al.}  [CLEO Collaboration],
  Phys.\ Rev.\  D {\bf 77}, 092003 (2008)
  [arXiv:0802.2664 [hep-ex]].

\bibitem{RosnerPP09}
  B.~Bhattacharya and J.~L.~Rosner,
  arXiv:0911.2812 [hep-ph].

\bibitem{CC89} L.L. Chau and H.Y. Cheng, Phys. Rev. D {\bf 39}, 2788 (1989);
L.L. Chau, H.Y. Cheng, and T. Huang, Z. Phys. C {\bf 53}, 413 (1992).

\bibitem{Li} X.Y. Li and S.F. Tuan, DESY Report No. 83-078
(unpublished); X.Y. Li, X.Q. Li, and P. Wang, Nuovo Ciemento
{\bf 100A}, 693 (1988).

\bibitem{KLOE}
  F.~Ambrosino {\it et al.},
  JHEP {\bf 0907}, 105 (2009)
  [arXiv:0906.3819 [hep-ph]].

\bibitem{Chau91} L.L. Chau, H.Y. Cheng, W.K. Sze, H. Yao, and B. Tseng,
Phys. Rev. D {\bf 43}, 2176 (1991).

\bibitem{FKS} T. Feldmann, P. Kroll, and B. Stech, Phys. Rev. D {\bf 58}, 114006 (1998); \pl {\bf B449}, 339 (1999).

\bibitem{Lipkin80} H.J. Lipkin, \prl {\bf 44}, 710 (1980).

\bibitem{CKMfitter} CKMfitter Group, J. Charles {\it et al.,} Eur.
Phys. J. C {\bf 41}, 1 (2005) and updated results from
http://ckmfitter.in2p3.fr.

\bibitem{UTfit} UTfit Collaboration, M. Bona {\it et
al.,} JHEP {\bf 0507}, 028 (2005) and updated results from
http://utfit.roma1.infn.it.

\bibitem{BelleDs}
  E.~Won, B.~R.~Ko {\it et al.} [Belle Collaboration],
  arXiv:0910.3052 [hep-ex].

\bibitem{Bigi}
  I.~I.~Y.~Bigi and H.~Yamamoto,
  Phys.\ Lett.\  B {\bf 349}, 363 (1995)
  [arXiv:hep-ph/9502238].

\bibitem{Gao}
  D.~N.~Gao,
  Phys.\ Lett.\  B {\bf 645}, 59 (2007)
  [arXiv:hep-ph/0610389].

\bibitem{CLEO:RD0}
  Q.~He {\it et al.}  [CLEO Collaboration],
  Phys.\ Rev.\ Lett.\  {\bf 100}, 091801 (2008)
  [arXiv:0711.1463 [hep-ex]].

\bibitem{PDG}
C.~Amsler {\it et al.}  [Particle Data Group], Phys.\ Lett.\  B {\bf 667}, 1 (2008).


\bibitem{CLEO:DsKKpi}
  R.~E.~Mitchell {\it et al.}  [CLEO Collaboration],
  Phys.\ Rev.\  D {\bf 79}, 072008 (2009)
  [arXiv:0903.1301 [hep-ex]].

\bibitem{CLEOrhoeta}
P. Naik et al. [CLEO Collaboration],
  arXiv:0910.3602 [hep-ex].

\bibitem{CLEOomega}
  J.~Y.~Ge {\it et al.}  [CLEO Collaboration],
  Phys.\ Rev.\  D {\bf 80}, 051102 (2009)
  [arXiv:0906.2138 [hep-ex]].

\bibitem{Kass} R. Kass, talk presented at 2009 Europhysics Conference on High Energy Physics, July 16-22, 2009, Krakow, Poland.

\bibitem{Chiang04}
  C.~W.~Chiang, M.~Gronau, Z.~Luo, J.~L.~Rosner and D.~A.~Suprun,
  Phys.\ Rev.\  D {\bf 69}, 034001 (2004)
  [arXiv:hep-ph/0307395].

\bibitem{BaBar:piphi}
  B.~Aubert {\it et al.}  [BABAR Collaboration],
  Phys.\ Rev.\  D {\bf 76}, 011102 (2007)
  [arXiv:0704.3593 [hep-ex]].

\bibitem{CLEO:piphi}
  C.~Cawlfield {\it et al.}  [CLEO Collaboration],
  Phys.\ Rev.\  D {\bf 74}, 031108 (2006)
  [arXiv:hep-ex/0606045].

\bibitem{Belle:piphi}
  K.~Abe {\it et al.}  [Belle Collaboration],
  Phys.\ Rev.\ Lett.\  {\bf 92}, 101803 (2004)
  [arXiv:hep-ex/0308037].

\bibitem{E791}  E.M. Aitala {\it et al.} [E791 Collaboration],
Phys. Rev. Lett.  {\bf 86}, 765 (2001).

\bibitem{BaBarrhopi}
  B.~Aubert {\it et al.}  [BABAR Collaboration],
  Phys.\ Rev.\  D {\bf 79}, 032003 (2009)
  [arXiv:0808.0971 [hep-ex]].

\bibitem{KLOE:Ds3pi}
  J.~M.~Link {\it et al.}  [FOCUS Collaboration],
  Phys.\ Lett.\  B {\bf 585}, 200 (2004)
  [arXiv:hep-ex/0312040].

\bibitem{CLEO1989}
  W.Y. Chen {\it et al.}  [CLEO Collaboration],
  Phys.\ Lett. \  B {\bf 226}, 192 (1989).

\bibitem{BSW} M. Wirbel, B. Stech, and M. Bauer, Z. Phys. C {\bf 29}, 637
(1985); M. Bauer, B. Stech, and M. Wirbel, ibid. C {\bf 34},
103 (1987).

\bibitem{Buras96} G. Buchalla, A.J. Buras, and M.E. Lautenbacher,
Rev. Mod. Phys. {\bf 68}, 1125 (1996).

\bibitem{Melikhov}
  D.~Melikhov and B.~Stech,
  Phys.\ Rev.\  D {\bf 62}, 014006 (2000)
  [arXiv:hep-ph/0001113].

\bibitem{CLEO:FF}
  D.~Besson {\it et al.}  [CLEO Collaboration],
  Phys.\ Rev.\  D {\bf 80}, 032005 (2009)
  [arXiv:0906.2983 [hep-ex]].

\bibitem{BallfV} P. Ball, G.W. Jones, and R. Zwicky, Phys. Rev. D {\bf 75}, 054004 (2007).

\bibitem{CC94} L.L. Chau and H.Y. Cheng, \pl {\bf B333}, 514 (1994).

\bibitem{BBNS} M. Beneke, G. Buchalla, M. Neubert, and C.T. Sachrajda,
Phys. Rev. Lett. {\bf 83}, 1914 (1999); Nucl. Phys. B {\bf 591}, 313 (2000).

\bibitem{Lai}
  J.~H.~Lai and K.~C.~Yang,
  Phys.\ Rev.\  D {\bf 72}, 096001 (2005)
  [arXiv:hep-ph/0509092].


\bibitem{Donoghue} J.F. Donoghue, Phys. Rev. D {\bf 33}, 1516 (1986).

\bibitem{Chenga1a2} H.Y. Cheng, Eur. Phys. J. C {\bf 26}, 551 (2003).

\bibitem{Zen} P. \.Zenczykowski, Acta Phys. Polon. B {\bf 28}, 1605
(1997) [hep-ph/9601265].

\bibitem{Weinberg} S. Weinberg, {\it The Quantum Theory of Fields,
Volume I} (Cambridge, 1995), Sec. 3.8.

\bibitem{GronauD} M. Gronau, Phys. Rev. Lett. {\bf 83}, 4005 (1999).

\bibitem{Chen} Y. Chen, A. Alexandru, S.J. Dong, T. Draper, Horv\'ath,
F.X. Lee, K.F. Liu, N. Mathur, C. Morningstar, M. Peardon, S.
Tamhankar, B.L. Yang, and J.B. Zhang, Phys. Rev. D {\bf 73}, 014516 (2006).

\bibitem{CCL} H.Y. Cheng, C.K. Chua, and K.F. Liu, Phys. Rev. D {\bf 74}, 094005 (2006).

\bibitem{BESgammapipi}
  M.~Ablikim {\it et al.},
  Phys.\ Lett.\  B {\bf 642}, 441 (2006)
  [arXiv:hep-ex/0603048].

\bibitem{Carlson}
  C.~E.~Carlson, J.~J.~Coyne, P.~M.~Fishbane, F.~Gross and S.~Meshkov,
  Phys.\ Lett.\  B {\bf 99}, 353 (1981).

\bibitem{Cornwall}
  J.~M.~Cornwall and A.~Soni,
  Phys.\ Rev.\  D {\bf 29}, 1424 (1984);
    Phys.\ Rev.\  D {\bf 32}, 764 (1985).


\bibitem{Chanowitz} M.S. Chanowitz, Phys.\ Rev.\ Lett.\ {\bf 95}, 172001 (2005).

\bibitem{Chao} K.T. Chao, X.G. He, and J.P. Ma, Eur. Phys. J. C {\bf 55}, 417 (2008);  Phys.\ Rev.\ Lett.\  {\bf 98}, 149103 (2007).

\bibitem{Pham} X.Y. Pham, Phys. Lett. B {\bf 193}, 331 (1987).

\bibitem{Lipkin} H.J. Lipkin, \pl {\bf B283}, 412 (1992); in {\it
Proceedings of the 2nd International Conference on $B$ Physics and
$CP$ Violation}, Honolulu, Hawaii, 1997, edited by T.E. Browder
{\it et al.} (World Scientific, Singapore, 1998), p.436.

\bibitem{Kamal}
  A.~N.~Kamal, N.~Sinha and R.~Sinha,
  Phys.\ Rev.\  D {\bf 39}, 3503 (1989).

\bibitem{Fajfer}
  S.~Fajfer, A.~Prapotnik, P.~Singer and J.~Zupan,
  Phys.\ Rev.\  D {\bf 68}, 094012 (2003)
  [arXiv:hep-ph/0308100].

\bibitem{Gronau2}
  M.~Gronau and J.~L.~Rosner,
  Phys.\ Rev.\  D {\bf 79}, 074022 (2009)
  [arXiv:0903.2287 [hep-ph]].

\bibitem{Gronau}
  M.~Gronau and J.~L.~Rosner,
  Phys.\ Rev.\  D {\bf 79}, 074006 (2009)
  [arXiv:0902.1363 [hep-ph]].

\bibitem{Benayoun:1999fv}
M.~Benayoun, L.~DelBuono, S.~Eidelman, V.~N.~Ivanchenko and
H.~B.~O'Connell,
  Phys.\ Rev.\  D {\bf 59}, 114027 (1999)
  [arXiv:hep-ph/9902326];
  A.~Kucukarslan and U.~G.~Meissner,
  Mod.\ Phys.\ Lett.\  A {\bf 21}, 1423 (2006)
  [arXiv:hep-ph/0603061];
   M.~Benayoun, P.~David, L.~DelBuono, O.~Leitner and H.~B.~O'Connell,
  Eur.\ Phys.\ J.\  C {\bf 55}, 199 (2008)
  [arXiv:0711.4482 [hep-ph]];
  W.~Qian and B.~Q.~Ma,
  Phys.\ Rev.\  D {\bf 78}, 074002 (2008).

\bibitem{CCSfsi} H.Y.~Cheng, C.K.~Chua, and A.~Soni, Phys. Rev. D {\bf
71}, 014030 (2005).

\bibitem{CLEO:omega}
  J.~Y.~Ge {\it et al.}  [CLEO Collaboration],
  Phys.\ Rev.\  D {\bf 80}, 051102 (2009)
  [arXiv:0906.2138 [hep-ex]].

\bibitem{MarkIII}
D.M. Coffman {\it et al.} [Mark III Collaboration], Phys. Rev.
D {\bf 45} 2196 (1992).

\bibitem{Kamal99}
  El Hassan El Aaoud and A.~N.~Kamal,
  Phys.\ Rev.\  D {\bf 59}, 114013 (1999)
  [arXiv:hep-ph/9910350].

\bibitem{ChengSP}
  H.~Y.~Cheng,
  Phys.\ Rev.\  D {\bf 67}, 034024 (2003)
  [arXiv:hep-ph/0212117].

\bibitem{ChengAP}
  H.~Y.~Cheng,
  Phys.\ Rev.\  D {\bf 67}, 094007 (2003)
  [arXiv:hep-ph/0301198].

\bibitem{ChengTP}
  H.~Y.~Cheng,
  Phys.\ Rev.\  D {\bf 68}, 014015 (2003)
  [arXiv:hep-ph/0303195].

\end{thebibliography}

\end{document}